\documentclass[11pt,a4paper,reqno]{amsart}

\usepackage[utf8]{inputenc}
\usepackage[T1]{fontenc}
\usepackage{fouriernc}
 \usepackage{extarrows}

\def\gbf#1{\mbox{\boldmath$#1$}} 
 
 \numberwithin{equation}{section}

 \usepackage[pdftex,
                paper=a4paper,
                portrait=true,
                textwidth=165mm,
                textheight=238mm,
                tmargin=2.5cm,
                marginratio=1:1]{geometry}

\theoremstyle{definition}

\theoremstyle{remark}

 \newtheorem*{note}{Note added}

\newtheorem*{acknowledgements}{Acknowledgements}

\newcommand{\CC}{\ensuremath{\mathbb{C}}}

\newcommand{\ZZ}{\ensuremath{\mathbb{Z}}}

\renewcommand\d{\operatorname{d \!}}

\newcommand\str{\operatorname{str}}

\newcommand\tr{\operatorname{tr}}

 \newcommand{\coh}{\operatorname{coh}}

\newcommand{\End}{\operatorname{End}}

\newcommand{\Ho}{\operatorname{Ho}}  
\newcommand{\Hom}{\operatorname{Hom}}

\newcommand{\Rep}{\operatorname{Rep}}

\title{Boundary coupling of Lie algebroid Poisson sigma models and representations up to homotopy}

\author{Alexander Quintero V\'{e}lez} 
  
\address{Department of Mathematics\\ University of Glasgow\\ Glasgow\\ G12 8QW\\ United Kingdom}
\email{Alexander.QuinteroVelez@glasgow.ac.uk}

\begin{document}

\begin{abstract}
A general form for the boundary coupling of a Lie algebroid Poisson sigma model is proposed. The approach involves using the Batalin-Vilkovisky formalism in the AKSZ geometrical version, to write a BRST-invariant coupling for a representation up to homotopy of the target Lie algebroid or its subalgebroids. These considerations lead to a conjectural description of topological D-branes on generalized complex manifolds, which includes A-branes and B-branes as special cases. 
\end{abstract}

 \maketitle

\section{Introduction}
A Poisson sigma model \cite{Ik94,SS94} is a two-dimensional topological field theory with a Poisson mani\-fold for target space. These models are closely related to other theories of interest such as two-dimensional gravity, two-dimensional Yang-Mills theory and gauged WZW models. They were also the object of much recent attention due to their connection with the program of deformation quantization of Poisson manifolds \cite{CF00}. 

In this paper we will consider a variant of the usual Poisson sigma model in which the target space geometry consists of a Lie algebroid $E$ over a base manifold $X$. These theories were introduced by Bonechi and Zabzine \cite{BZ07} in order to give a convenient description of the moduli space of Lie algebroid morphisms modulo gauge transformations. They have also been worked out in \cite{CQZ10} by employing the Batalin-Vilkovisky algorithm in the AKSZ formulation. We will refer to them as Lie algebroid Poisson sigma models. 

An important aspect of Lie algebroid Poisson sigma models is that they admit nontrivial boundary conditions. Indeed, as pointed out in \cite{BZ07,CQZ10} and further discussed below, these boundary conditions are set by specifying a Lie sublagebroid $F$ of $E$ over a submanifold $Y \subset X$. An analogous result was first obtained in the Poisson sigma model context by Cattaneo and Felder in \cite{CF04} (see also \cite{CalFal04,BZ05}).

The purpose of this paper is to find more general boundary conditions for a Lie algebroid Poisson sigma model by allowing interactions between bulk and boundary degrees of freedom. To this aim, we follow the strategy that has been used to classify B-type boundary conditions in nonlinear sigma models and Landau-Ginzburg models \cite{Diac01,KL03,BHLS06,Laz05,HHP08,Bergman08}. It turns out that the most general boun\-dary interaction corresponds to a $\ZZ$-graded vector bundle on $Y$ equipped with an $F$-superconnection. Moreover, BRST invariance requires the $F$-superconnection to be flat. In the terminology of Arias Abad and Crainic \cite{AAC09}, such a flat $F$-superconnection defines a {\it representation up to homotopy} of $F$. When combined with a careful study of the structure of the boundary observables, this fact leads to the identification of boundary conditions for a Lie algebroid Poisson sigma model with objects in the DG-category of representations up to homotopy of $F$ (or rather its homotopy category). 

One of the main reasons for the interest in this problem is the light that it may shed on the geo\-metry of D-branes in topologically twisted nonlinear sigma models. In \cite{Kap04} (see also \cite{KL05,KL07}), it was found that the geometry of such D-branes can be conveniently formulated in terms of generalized complex structures \cite{Hitchin03,Gualt03,Cav04}. Since a generalized complex structure on a manifold $X$ gives rise to a complex Lie algebroid $E$, we can attach to it a Lie algebroid Poisson sigma model. An important observation made in \cite{CQZ10} is that the A-model (respectively, B-model) can be obtained by an appropriate gauge fixing of a Lie algebroid Poisson sigma model in which the underlying generalized complex manifold $X$ is a symplectic manifold (respectively, ordinary complex manifold). The above description then shows (or rather, suggests) that topological D-branes in the A- and B-model wrapped on a gene\-ralized complex submanifold $Y \subset X$ can naturally be thought of as representations up to homotopy of a Lie algebroid canonically associated with $Y$. In the former case, this interpretation enables us to identify topological D-branes in the A-model with complexes of higher rank coisotropic A-branes, exac\-tly as originally argued by Herbst in \cite{Herbst10}. (We are here ignoring the perturbative effects that bring noncommutativity into the A-model.) In the latter case, we will recover the familiar corres\-pondence between topological D-branes in the B-model and objects of the bounded derived category of coherent sheaves on $X$. Thus, what we learn is that the notion of representation up to homotopy seems to be the correct framework to understand D-branes in general topologically twisted nonlinear sigma models. 

The outline of the paper is as follows. We begin with a review of the basic definitions and results concerning representations up to homotopy of Lie algebroids in section~\ref{Sect:2}. In section~\ref{Sect:3}, we give a brief presentation of Lie algebroid Poisson sigma models and their basic relevant properties. We then go on to discuss the allowed boundary conditions for these models in section~\ref{Sect:BC}. Section~\ref{Sect:AKSZ} contains an account of the Batalin-Vilkovisky formulation of Lie algebroid Poisson sigma models. We follow this with a discussion of the BRST-invariant boundary observables in section~\ref{Sect:6}. In section~\ref{Sect:7}, we couple the Lie algebroid Poisson sigma model of section~\ref{Sect:AKSZ} to an ordinary $F$-connection. We show that BRST invariance of the boundary interaction amounts to the condition that the $F$-connection be flat, that is to say, a representation of $F$. Thereupon, in section~\ref{Sect:8}, we explain how the boundary observables are modified when the model is coupled to such a representation. After this warm-up we turn to $F$-superconnections in section~\ref{Sect:9}. We show that the background $F$-superconnections to which the Lie algebroid Poisson sigma model can be coupled are precisely those for which the curvature vanishes, in other words those that define representation up to homotopy of $F$. The effect on the boundary observables when coupling to a representation up to homotopy is considered in section~\ref{Sect:10}. Finally, in section~\ref{Sect:11}, we offer our conclusions and thoughts in connection with topological D-branes.

\begin{note}
As pointed out to us by one of the referees, many of the calculations in this paper can be most conveniently and elegantly done using the language of NQ-manifolds espoused in \cite{BQZ11}.
\end{note}

\section{Representations up to homotopy} 
\label{Sect:2}
In this section we review the definition and basic facts regarding representations up to homotopy of Lie algebroids, emphasizing those points most relevant to the present work. Our presentation is slightly different, but equivalent to, that in the original paper by Arias Abad and Crainic \cite{AAC09}. 

First let us recall the necessary definitions. A {\it Lie algebroid} is a vector bundle $E$ over a smooth manifold $X$ together with a Lie algebra structure $[,]$ on the space of sections of $E$ and a bundle map $\rho\colon E \to X$, called the anchor, satisfying the compatibility condition
$$
[s_1,fs_2]=f[s_1,s_2]+\rho(s_1)(f) s_2,
$$ 
for any two sections $s_1$, $s_2$ of $E$ and any function $f$ on $X$. If one fixes local coordinates $\{x^{\mu} \}$ over a trivializing neighborhood of $X$ where $E$ admits a basis of local sections $\{e_I \}$, we have structure functions $\rho^{\mu}_I(x^{\mu})$ and $C^K_{IJ}(x^{\mu})$ defined by
$$
\rho(e_I)=\rho^{\mu}_I(x^{\mu})\frac{\partial}{\partial x^{\mu}}, \qquad [e_I,e_J]=C^K_{IJ}(x^{\mu})e_K,
$$
where the summation over repeated indices is understood. The compatibility condition for a Lie algebroid translates into certain PDEs involving the structure functions.  

Before proceeding, it will be helpful to introduce a bit of notation. Given a Lie algebroid $E \to X$, we write $\Omega^{\bullet}(E)=\bigoplus_{k\geq 0}\Omega^{k}(E)$ to denote the space of sections of the exterior algebra bundle $\Lambda^{\bullet} E^*=\bigoplus_{k\geq 0}\Lambda^k E^*$. Then $\Omega^{\bullet}(E)$ is a graded module for the algebra $\Omega^0(X)$ of smooth functions on $X$. Its elements will be called $E${\it-forms}. If $V \to X$ is any vector bundle, we let $\Omega^{\bullet}(E,V)=\Omega^{\bullet}(E)\otimes_{\Omega^0(X)}\Omega^0(X,V)$ denote the $\Omega^{\bullet}(E)$-module of $E$-forms with values in $V$, where $\Omega^0(X,V)$ is the space of sections of $V$. 

The simplest Lie algebroid over $X$ is the tangent bundle $TX$ itself, with the usual bracket of vector fields and the identity as anchor map.  All the usual constructions of differential geometry make sense when one replaces $TX$ with an arbitrary Lie algebroid $E$. For example, on $\Omega^{\bullet}(E)$ there is a degree $1$ differential $\mathrm{d}_E$ which squares to zero, analogous to the de Rham differential. It is defined using the obvious generalization of the Cartan formula to the Lie algebroid case, namely
\begin{align*}
\mathrm{d}_E \omega (s_1,\ldots,s_{k+1})&=\sum_{i=1}^{k+1}(-1)^{i-1}\rho(s_i)( \omega(s_1,\ldots,\widehat{s_i},\ldots,s_{k+1})) \\
&\qquad\qquad+\sum_{1\leq i<j\leq k+1}(-1)^{i+j}\omega([s_i,s_j],s_1,\ldots,\widehat{s_i},\ldots,\widehat{s_j},\ldots,s_{k+1}),
\end{align*}
where $\omega$ is an $E$-form of degree $k$ (an $E$-$k$-form for short) and $s_1,\ldots,s_{k+1}$ are sections of $E$. The cohomology associated with $\mathrm{d}_E$ is called the {\it Lie algebroid cohomology} of $E$ (with trivial coefficients). 

One can also consider a notion of connection on a Lie algebroid $E$ which is a natural extension of the usual concept of covariant connection. An $E${\it-connection} on a vector bundle $V$ over $X$ is a bilinear mapping $\nabla\colon \Omega^0(X,E)\times \Omega^0(X,V)\to \Omega^0(X,V)$, which satisfies the following conditions:
$$
\nabla_{fs}\phi=f\nabla_s \phi, \qquad \nabla_s(f \phi)=f \nabla_s \phi + \rho(s)(f) \phi,
$$
for any $f$ in $\Omega^0(X)$,  $\phi$ in $\Omega^0(X,V)$ and $s$ in $\Omega^0(X,E)$. The $E${\it-curvature} of $\nabla$ is the element $F_{\nabla}$ in $\Omega^2(E,\End(V))$ given by
$$
F_{\nabla}(s_1,s_2)(\phi)=\nabla_{s_1}\nabla_{s_2}\phi-\nabla_{s_2}\nabla_{s_1}\phi-\nabla_{[s_1,s_2]}\phi,
$$
for any $\phi$ in $\Omega^0(X,V)$ and $s_1,s_2$ in $\Omega^0(X,E)$. An $E$-connection $\nabla$ is said the be {\it flat} if its $E$-curvature is zero, i.e. $F_{\nabla}=0$. A flat $E$-connection on a vector bundle $V$ is sometimes called a {\it Lie algebroid representation} of $E$. 

Local expressions for $E$-connections can be obtained in a way similar to the covariant case. Let $\{ x^{\mu}\}$ be local coordinates in a trivializing neighborhood of $X$ for $E$ and $V$, and let $\{e_I\}$ and $\{v_a\}$ be the respective basis of local sections. Then we can define the $E${\it-connection coefficients} $A_{Ib}^a$ by
$$
\nabla_{e_I}v_b=A_{Ib}^a v_a,
$$ 
where repeated indices are summed over. If $\{v^a\}$ is the basis of local sections of $V^*$ dual to $\{v_a\}$, we should think of $A_I=A_{Ib}^a v^b \otimes v_a$ as components of an $E$-one-form with values in $\End(V)$. For a section $\phi=\phi^a v_a$ of $V$, one finds
$$
\nabla_{e_I}\phi=\big(\rho_I^{\mu}\partial_{\mu} \phi^a+A_{Ib}^a \phi^b\big)v_a,
$$
where we have written $\partial_{\mu}$ for $\partial/\partial x^{\mu}$. Thus the components of $\nabla_{e_I}\phi$ with respect to the basis of sections $\{v_a \}$, denoted $\nabla_I \phi^a$, are
$$
\nabla_I \phi^a = \rho_I^{\mu}\partial_{\mu} \phi^a+A_{Ib}^a \phi^b.
$$
By a familiar abuse of language, we shall refer to these as the {\it covariant derivative} of the section $\phi$. As an aside, we remark that the definition of the covariant derivative can be extended to any section of the bundle $V^{*\otimes k}\otimes V^{\otimes l}$ for $k,l \geq 0$.

The $E$-curvature $F_{\nabla}$ is also easily expressible in terms of the $E$-connection coefficients. If we write $F_{\nabla}(e_I,e_J)(v_b)=F_{IJ b}^av_a$, a direct computation yields
$$
F_{IJ b}^a=\rho^{\mu}_I \partial_{\mu}A_{Jb}^a-\rho^{\mu}_J \partial_{\mu}A_{Ib}^a+A_{Ic}^aA_{Jb}^c-A_{Jc}^aA_{Ib}^c-C^K_{IJ}A_{Kb}^a.
$$ 
This can be usefully rewritten in terms of the $\End(V)$-valued components $F_{IJ}=F_{IJ b}^a v^b \otimes v_a$ as
$$
F_{IJ}=\rho^{\mu}_I \partial_{\mu}A_{J}-\rho^{\mu}_J \partial_{\mu}A_{I}+[A_{I},A_{J}]-C^K_{IJ}A_{K},
$$
where the bracket $[,]$ denotes the usual commutator in $\End(V)$. Thus, if $\{ e^I \}$ is the basis of sections for $E^*$ dual to $\{ e_I \}$, then $F_{\nabla}=\frac{1}{2}F_{IJ}e^I \wedge e^J$.

Given a representation $V$, determined by a flat $E$-connection $\nabla$, one has an induced operator $\mathrm{d}_{E,\nabla}$ on $\Omega^{\bullet}(E,V)$, of degree $1$, whose definition is in the same spirit as the formula for the exterior derivative of a scalar-valued $E$-form:
\begin{align*}
\mathrm{d}_{E,\nabla} \omega (s_1,\ldots,s_{k+1})&=\sum_{i=1}^{k+1}(-1)^{i-1}\nabla_{s_i}( \omega(s_1,\ldots,\widehat{s_i},\ldots,s_{k+1})) \\
&\qquad\qquad+\sum_{1\leq i<j\leq k+1}(-1)^{i+j}\omega([s_i,s_j],s_1,\ldots,\widehat{s_i},\ldots,\widehat{s_j},\ldots,s_{k+1}).
\end{align*}
Here $\omega$ is a $V$-valued $E$-$k$-form and $s_1,\ldots,s_{k+1}$ are sections of $E$. It is a simple matter to check that $\mathrm{d}_{E,\nabla}^{2}=0$. The resulting cohomology is called the {\it Lie algebroid cohomology of $E$ with coefficients in} $V$. From the formula, it also follows that, if $\omega=\frac{1}{k!}\omega_{I_1 \cdots I_k}e^{I_1}\wedge \cdots \wedge e^{I_k}$ is a local expression for the $V$-valued $E$-$k$-form $\omega$, then the components of $\mathrm{d}_{E,\nabla} \omega$ are
$$
\left( \mathrm{d}_{E,\nabla} \omega \right)_{I_1 \cdots I_{k+1}}=\sum_{i=1}^{k+1}(-1)^{i-1}\nabla_{I_i} \omega_{I_1 \cdots \widehat{I_i} \cdots I_{k+1}} + \sum_{1\leq i<j\leq k+1}(-1)^{i+j}C_{I_i I_j}^J \omega_{J I_1 \cdots \widehat{I_i} \cdots \widehat{I_j} \cdots I_{k+1}}.
$$

Now we come to the definition of representation up to homotopy of a Lie algebroid $E\to X$.  We will formulate this in a way that exhibits the relation to Quillen's notion of superconnection. Let $V=\bigoplus_{k \in \ZZ}V^k$ be a $\ZZ$-graded vector bundle over the manifold $X$. We consider the space of $V$-valued $E$-forms $\Omega^{\bullet}(E,V)$ to be $\ZZ$-graded with respect to the total degree. An $E${\it-superconnection} on $V$ is an operator $D$ on $\Omega^{\bullet}(E,V)$ of degree $1$ which satisfies the Leibniz rule
$$
D(\sigma \wedge \omega)=\mathrm{d}_{E}\sigma \wedge \omega + (-1)^k \sigma \wedge D\omega,
$$
for any $\sigma$ in $\Omega^k(E)$ and $\omega$ in $\Omega^{\bullet}(E,V)$. The {\it curvature} of $D$ is defined as $D^2$, a $\Omega^{\bullet}(E)$-linear operator on $\Omega^{\bullet}(E,V)$ of degree $2$ which is given by multiplication by an element of $\Omega^{\bullet}(E,\End(V))$ of degree $2$ (cf.~Lemma~2.14 of \cite{AAC09}). If $D^2=0$, then we say that $D$ is a {\it flat} $E$-superconnection. By a {\it representation up to homotopy} of $E$ we mean a $\ZZ$-graded vector bundle $V$ endowed with a flat $E$-superconnection $D$. 

Before proceeding further, it is worth recalling that the space $\Omega^{\bullet}(E,\End(V))$ is naturally equipped with a $\ZZ$-graded algebra structure. The multiplication $\omega \wedge \eta$ of a $\Hom(V^{\bullet},V^{\bullet+p})$-valued $E$-$k$-form $\omega$ and a $\Hom(V^{\bullet},V^{\bullet+q})$-valued $E$-$l$-form $\eta$ is defined to be the $\Hom(V^{\bullet},V^{\bullet+p+q})$-valued $E$-$(k+l)$-form having components
$$
(\omega \wedge \eta)_{I_1\cdots I_{k+l}}=\frac{(k+l)!}{k! l!}(-1)^{p l}\omega_{I_1 \cdots I_k} \eta_{I_{k+1}\cdots I_{k+l}},
$$
where $\omega_{I_1 \cdots I_k}$ and $\eta_{I_{k+1}\cdots I_{k+l}}$ are the components of $\omega$ and $\eta$, respectively. Accordingly, we define the graded commutator of $\omega$ and $\eta$ by
$$
[\omega,\eta]=\omega \wedge \eta -(-1)^{(k+p)(l+q)}\eta\wedge \omega.
$$
Equipped with this commutator, $\Omega^{\bullet}(E,\End(V))$ becomes a $\ZZ$-graded Lie algebra.

Intuitively, one expects that a representation up to homotopy of $E$ is a $\ZZ$-graded vector bundle together with an $E$-connection which is ``flat up to homotopy''. To put this more precisely, suppose that $D$ is a representation up to homotopy on a $\ZZ$-graded vector bundle $V$. The Leibniz rule means that $D$ is entirely determined by its restriction to $\Omega^{\bullet}(E)$. One can then decompose
$$
D=\sum_{k \geq 0}D^{(k)},
$$
where $D^{(k)}$ is of partial degree $k$ with respect to the $\ZZ$-grading on $\Omega^{\bullet}(E)$. Clearly each $D^{(k)}$ for $k\neq 1$ is $\Omega^{\bullet}(E)$-linear and therefore it is given by multiplication by an element $\Omega^{(k)}$ of $\Omega^{k}(E,\Hom(V^{\bullet},V^{\bullet+1-k}))$. On the other hand, $D^{(1)}$ satisfies the Leibniz rule on each of the vector bundles $V^k$, so it must be of the form $\mathrm{d}_{E,\nabla}$, where $\nabla$ is an $E$-connection on $V$ which preserves the $\ZZ$-grading (cf.~Proposition~2.3~of~\cite{AAC09}). We can thus write, for $\omega$ in $\Omega^{\bullet}(E,V)$,
\begin{equation}
\label{eq:2.1}
D \omega=v \wedge \omega + \mathrm{d}_{E,\nabla}\omega +\sum_{k\geq 2}\Omega^{(k)}\wedge \omega,
\end{equation}
where we have put $v = \Omega^{(0)}$. From this formula, it is straightforward to show that the flatness condition becomes equivalent to
\begin{align}
\label{eq:2.2}
\begin{split}
& v^2=0,\\
&\mathrm{d}_{E,\nabla} v=0,\\
& \big[ v , \Omega^{(2)} \big]+ F_{\nabla}=0,\\
\end{split}
 \end{align}
 and 
\begin{equation}
\label{eq:2.3}
\big[ v , \Omega^{(n+1)}\big] + \mathrm{d}_{E,\nabla} \Omega^{(n)}+\sum_{k=2}^{n-1} \Omega^{(k)} \wedge  \Omega^{(n+1-k)}=0,
\end{equation}
for each $n \geq 2$. In terms of components these conditions take the forms
\begin{align}
\label{eq:localRepsGraded1}
\begin{split}
& v^2=0,\\
&\nabla_I v=0, \\
& v\Omega^{(2)}_{IJ}+\Omega^{(2)}_{IJ}v +F_{IJ}=0,
 \end{split}
 \end{align}
and 
\begin{align}
\label{eq:localRepsGraded2}
\begin{split}
& (-1)^{n+1}v \Omega^{(n+1)}_{I_1\cdots I_{n+1}}+\Omega^{(n+1)}_{I_1\cdots I_{n+1}}v +\sum_{k=1}^{n+1}(-1)^{k-1}\nabla_{I_k}\Omega^{(n)}_{I_1\cdots \widehat{I}_k\cdots I_{n+1}} \\
&\quad \quad + \sum_{1\leq k < l \leq n+1}(-1)^{k+l}C^{J}_{I_kI_l}\Omega^{(n)}_{JI_1\cdots \widehat{I}_k\cdots \widehat{I}_l\cdots I_{n+1}}+\sum_{k=2}^{n-1}\binom{n+1}{k}(-1)^{(1-k)(n+1-k)}\Omega^{(k)}_{I_1\cdots I_k}\Omega^{(n+1-k)}_{I_{k+1}\cdots I_{n+1}} =0,
\end{split}
\end{align}
for each $n \geq 2$, whereby the covariant derivative is $\nabla_I=\rho_I^{\mu}\partial_{\mu}+[A_I,\,]$. One may view \eqref{eq:2.2}-\eqref{eq:2.3} as the defining relations for a representation up to homotopy; see Proposition~3.2 of \cite{AAC09}. The first identity in \eqref{eq:2.2} implies that we have a cochain complex of vector bundles
$$
\cdots \xlongrightarrow{v} V^{k-1} \xlongrightarrow{v} V^{k} \xlongrightarrow{v} V^{k+1} \xlongrightarrow{v} \cdots
$$
The second equation in \eqref{eq:2.2} express the fact that $v$ is covariantly constant with respect to the $E$-connection $\nabla$. The last equation in \eqref{eq:2.2} indicates that the $E$-connection $\nabla$ fails to be flat up to terms involving the homotopy $\Omega^{(2)}$ and the differential $v$.

We may restate the above more concisely as follows. In view of \eqref{eq:2.1}, the $F$-superconnection $D$ may be written locally as $D=\mathrm{d}_E+\alpha$, where $\alpha$ is the $\End(V)$-valued $E$-form defined by
\begin{equation}
\label{eq:2.6}
\alpha=v + A_I e^{I} + \sum_{k\geq 2}\frac{1}{k!}\Omega^{(k)}_{I_1\cdots I_k}e^{I_1}\wedge \cdots \wedge e^{I_k}.
\end{equation}
It is then elementary algebra to show that
$$
D^2 \omega =\left( \mathrm{d}_{E}\alpha+\frac{1}{2}[\alpha,\alpha] \right) \wedge\omega,
$$
for any $\omega$ in $\Omega^{\bullet}(E,\End(V))$. Thus, the totality of equations \eqref{eq:localRepsGraded1}-\eqref{eq:localRepsGraded2} is equivalent to the single statement that $\alpha$ satisfies the {\it Maurer-Cartan equation}
\begin{equation}
\label{eq:2.7}
 \mathrm{d}_{E}\alpha+\frac{1}{2}[\alpha,\alpha]=0.
\end{equation}

To conclude this section we note that all representations up to homotopy of a Lie algebroid $E$ can be naturally organized into a DG-category, which we denote by $\Rep(E)$. Its objects are, of course, representations up to homotopy of $E$, which we regard as pairs $(V,D)$. Given two such pairs $(V,D)$ and $(V',D')$ we define the space of morphisms to be $\ZZ$-graded vector space $\Omega^{\bullet}(E,\Hom(V,V'))$ with differential $\partial_{D,D'}$ acting on homogeneous elements of degree $k$ as
$$
\partial_{D,D'} \omega = D' \wedge \omega - (-1)^k \omega \wedge D.
$$
Locally, $\partial_{D,D'}$ may be expressed by
$$
\partial_{D,D'} \omega = \mathrm{d}_E \omega + \alpha' \wedge \omega - (-1)^k \omega \wedge \alpha,
$$
with $\alpha$ and $\alpha'$ as in the previous paragraph. In particular, if we take $D=D'$, then $\partial_{D,D}= \mathrm{d}_E + [\alpha,\,]$.

It is also worth noting that one can make an ordinary category out of $\Rep(E)$ by redefining morphisms between $(V,D)$ and $(V',D')$ to be the cohomology of $\partial_{D,D'}$. We call this category the {\it homotopy category} of $\Rep(E)$, and we denote it by $\Ho\Rep(E)$.\footnote{Strictly speaking, the homotopy category of $\Rep(E)$ is obtained by taking the degree zero cohomology of the morphism spaces in $\Rep(E)$. Nevertheless, in this paper we will keep morphisms of all degrees, since this conforms better to physical conventions.}

\section{Lie algebroid Poisson sigma models}
\label{Sect:3}
In \cite{BZ07} Bonechi and Zabzine introduced a two-dimensional topological field theory associated to any Lie algebroid, extending the well-known Poisson sigma model. We call this theory, in analogy with \cite{Zuc08}, a {\it Lie algebroid Poisson sigma model}. In this section, we will describe briefly the essential properties of these models. 

To begin with, we recall that a Poisson sigma model is specified by its target space, which is just a given Poisson manifold $X$. A field configuration is then simply a bundle map from the tangent space $T \Sigma$ of a two-dimensional oriented manifold $\Sigma$, possibly with boundary, to the cotangent bundle $T^*X$ of $X$. Such a map is given by a pair $(\phi,\eta)$ consisting of a base map $\phi\colon \Sigma \to X$ and a one-form $\eta$ on $\Sigma$ which takes values in $\phi^*T^*X$. Similarly, a Lie algebroid Poisson sigma model is specified by its target space and its target algebroid: the former is a given smooth manifold $X$, and the latter is a Lie algebroid $E \to X$ over $X$ with anchor map $\rho\colon E \to TX$ and Lie bracket $[,]$. A field configuration is then a pair $(\phi,\psi)$, where $\phi$ is again a map from $\Sigma$ to $X$ and $\psi$ is a one-form on $\Sigma$ taking values in $\phi^*E$. If $\{\sigma^{\alpha}\}$ and $\{x^{\mu}\}$ are respectively local coordinates in $\Sigma$ and $X$, and $\{e_I\}$ is a local trivialization for $E$, then $\phi$ can be described in terms of functions $\phi^{\mu}$ on $\Sigma$ and $\psi$ is specified by giving differential one-forms $\psi^I=\psi^I_{\alpha} \d \sigma^{\alpha}$. In addition to the field configuration $(\phi,\psi)$, we also require a bosonic field $\eta$, a one-form on $\Sigma$ with values in $\phi^*T^*X$, and a scalar field $\lambda$, a section of the pull-back bundle $\phi^*E^*$. Using the above coordinates, $\eta$ can be described locally via differential one-forms $\eta_{\mu}=\eta_{\mu \alpha}\d \sigma^{\alpha}$ and $\lambda$ is represented by functions $\lambda_I$. The action for the theory is
\begin{equation}
\label{eq:3.1}
S_0=\int_{\Sigma}\eta_{\mu} \wedge \d \phi^{\mu}+\psi^I \wedge \d \lambda_I + \rho^{\mu}_I\psi^I \wedge \eta_{\mu}+\frac{1}{2}C^K_{IJ}\lambda_K \psi^I \wedge \psi^J,
\end{equation}
where $\rho^{\mu}_I$ and $C^K_{IJ}$ denote respectively the structure functions of $\rho$ and $[,]$. As noted in \cite{BZ07}, this action can be directly interpreted as a Poisson sigma model with target $E^*$. (This merely corresponds to the fact that the dual bundle $E^*$ has a natural Poisson structure; see, for example, \cite{Fer02}.) It follows, then, that the equations of motion in the bulk describe Lie algebroid morphisms from $T\Sigma$ to $T^*E^*$.

Up to a boundary term, the action \eqref{eq:3.1} is invariant under the following infinitesimal gauge transformations
\begin{align}
\label{eq:3.2}
\begin{split}
\delta \phi^{\mu}&=\rho^{\mu}_I \xi^I, \\
\delta \lambda_I &= \rho^{\mu}_I \zeta_{\mu}+ C^K_{IJ}\lambda_K \xi^J, \\
\delta \eta_{\mu} &=-\d \zeta_{\mu} - \partial_{\mu}C^K_{IJ}\lambda_K \psi^I\xi^J-\partial_{\mu}\rho^{\nu}_I \psi^I \zeta_{\nu} - \partial_{\mu}\rho^{\nu}_I \eta_{\nu}\xi^I, \\
\delta \psi^I &=-\d \xi^I - C^I_{JK}\psi^J \xi^K,
\end{split}
\end{align}
where $\zeta$ is a section of $\phi^*T^*X$ and $\xi$ is a section of $\phi^*E$. In fact, under \eqref{eq:3.2}, the change in $S_0$ is
\begin{equation}
\label{eq:3.3}
\delta S_0=-\int_{\Sigma} \d \,(\zeta_{\mu} \d \phi^{\mu}+ \xi^I \d \lambda_I ).
\end{equation}
We also note that the commutator of two infinitesimal gauge transformations is a gauge transformation only {\it on-shell}; that is, we must use the on-shell equations of motion of \eqref{eq:3.1}. Quantization of the model is therefore more subtle, and requires the use of the Batalin-Vilkovisky formalism. Before launching into that technical discussion, let us first take a closer look at the boundary conditions for the fields and how they affect the gauge transformations.

\section{Boundary conditions}
\label{Sect:BC}
Boundary conditions for a Lie algebroid Poisson sigma model were first considered in \cite{BZ07}. Here we reproduce some of their results. In the discussion, we will assume for simplicity that the boundary $\partial \Sigma$ has a single connected component. We also assume that $\partial \Sigma$ is closed and is being parametrized with an angular variable $\tau$.  

The first constraint comes from the equation of motion for the fields $\phi$ and $\lambda$. In fact, under an infinitesimal change in $\phi$ and $\lambda$, the change in the action \eqref{eq:3.1} consists of a bulk term minus a boundary term:
\begin{align*}
\delta S_0&= \int_{\Sigma} \left( \d \eta_{\mu}+\partial_{\mu}\rho^{\nu}_I \psi^I \wedge \eta_{\nu}+\frac{1}{2}\partial_{\mu}C^K_{IJ}\lambda_K \psi^I\wedge \psi^J \right) \delta \phi^{\mu} \\
& \qquad+\left( \d \psi^I+\frac{1}{2}C^I_{JK} \psi^J\wedge \psi^K \right) \delta \lambda_I -\int_{\partial \Sigma} i^*(\eta_{\mu}\delta \phi^{\mu}+\psi^I \delta \lambda_I).
\end{align*}
Here $i\colon \partial \Sigma \to \Sigma$ denotes the inclusion of the boundary. We require both the bulk equation of motion and the integrand $i^*(\eta_{\mu}\delta \phi^{\mu}+\psi^I \delta \lambda_I)$ to vanish. This imposes the constraint on the boundary
\begin{equation}
\label{eq:4.1}
\eta_{\mu\tau}\delta \phi^{\mu}+\psi^I_{\tau}\delta \lambda_I=0,
\end{equation}
where $i^*\eta_{\mu}=\eta_{\mu\tau}\d \tau$ and $i^*\psi^I=\psi^I_{\tau}\d \tau$. Another constraint comes from requiring that the boundary term \eqref{eq:3.3}
 vanishes. This amounts to the following condition on the gauge parameters $\zeta$ and $\xi$:
 \begin{equation}
\label{eq:4.2}
\zeta_{\mu}\partial_{\tau}\phi^{\mu}+\xi^I\partial_{\tau}\lambda_I=0.
\end{equation}
Finally, one requires that these boundary conditions are invariant under the residual gauge transformations (that is, the gauge transformations \eqref{eq:3.1} restricted to the boundary $\partial \Sigma$). 

In \cite{BZ07} it was found that a wide class of boundary conditions that satisfy the previous requirements can be realized by Lie subalgebroids of $E$. To make this precise, let $F$ be a Lie subalgebroid of $E$ over a submanifold $Y \subset X$. We denote by $F^{\perp} \subset E^*$ the annihilator of $F$, and by $N^*Y$ the conormal bundle of $Y$. The first part of the boundary conditions requires that $\phi$ maps $\partial \Sigma$ to $Y$. This implies that the allowed changes $\delta \phi$ at the boundary $\partial \Sigma$ lie in $\phi^*TY$. We supplement this condition with the requirement that the pullback of $\psi$ to $\partial \Sigma$ takes values in $\phi^*F$. For equation \eqref{eq:4.1} to hold, therefore, we have to assume that at the boundary $\partial \Sigma$, $\eta$ takes values in $\phi^*N^*Y$ and $\lambda$ takes values in $\phi^*F^{\perp}$. In order to satisfy the other equation \eqref{eq:4.2}, we also need to assume that the gauge parameters $\zeta$ and $\xi$ are restricted to take values in $\phi^*N^*Y$ and $\phi^*F$, respectively.

We now assert that these boundary conditions are invariant under the residual gauge transformations. To see this, and for future purposes, it is useful to analyze the boundary conditions in local coordinates. We choose adapted local coordinates $\{x^{\mu'},x^{\mu''} \}$ on $X$ and a basis of sections $\{e_{I'},e_{I''}\}$ of $E$ so that the Lie algebroid structure over $F \to Y$, has coordinates $\{x^{\mu'}\}$ in the base, sections $\{e_{I'}\}$, anchor functions $\rho^{\mu}_{I'}(x^{\mu'},0)$ and Lie algebra structure functions $C^{K'}_{I' J'}(x^{\mu'},0)$. In these coordinates the boundary conditions are 
\begin{equation*}
\phi^{\mu''}=0,\qquad\lambda_{I'}=0,\qquad\eta_{\mu' \tau}=0,\qquad\psi^{I''}_{\tau}=0,\qquad\zeta_{\mu'}=0,\qquad \xi^{I''}=0.
\end{equation*}
Hence, the requirement that the gauge transformations restricted to the boundary should leave the boundary conditions invariant leads to the equations
\begin{align*}
0=&\rho^{\mu''}_{I'}(\phi^{\mu'},0) \xi^{I'}, \\
0=&\rho^{\mu''}_{I'}(\phi^{\mu'},0) \zeta_{\mu''}+ C^{K''}_{I'J'}(\phi^{\mu'},0)\lambda_{K''} \xi^{J'}, \\
0=&-\partial_{\mu'}C^{K''}_{I'J'}(\phi^{\mu'},0)\lambda_{K''} \psi^{I'}_{\tau}\xi^{J'} -\partial_{\mu'}\rho^{\mu''}_{I'}(\phi^{\mu'},0) \psi^{I'}_{\tau} \zeta_{\mu''}- \partial_{\mu'}\rho^{\mu''}_{I'}(\phi^{\mu'},0) \eta_{\mu''\tau}\xi^{I'}, \\
0=&-C^{I''}_{J'K'}(\phi^{\mu'},0)\psi^{J'}_{\tau} \xi^{K'}.
\end{align*}
These are satisfied by virtue of the fact that, with respect to the coordinates above, $\rho^{\mu''}_{I'}(x^{\mu'},0)=0$ and $C^{K''}_{I'J'}(x^{\mu'},0)=0$; see section~2.1 of  \cite{BZ07}. 

In summary, we have found that a boundary condition in a Lie algebroid Poisson sigma model corresponds to a choice of a Lie subalgebroid $F$ of the target algebroid $E$. We should point out that this result can alternatively be deduced more directly from the arguments of \cite{CF04}, if one notices that $F^{\perp} \subset E^*$ is a coisotropic submanifold. Finally, let us remark that a classical solution of the equations of motion with this boundary condition is given by a Lie algebroid morphism $T\Sigma \to T^*E^*$ such that its restriction to the boundary is a Lie algebroid morphism from $T\partial \Sigma$ to $N^*F^{\perp}$.

\section{Batalin-Vilkovisky formulation}
\label{Sect:AKSZ}
In this section, we will briefly sketch the relevant conceptual and computational features of the Batalin-Vilkovisky treatment of Lie algebroid Poisson sigma models. For a detailed description, the reader should consult \cite{CQZ10}.

Suppose that we are given the data of section~\ref{Sect:3}, namely, a two-dimensional oriented manifold $\Sigma$ and a Lie algebroid $E \to X$ over a smooth manifold $X$. In order to write down the Batalin-Vilkovisky action functional we shall make use of the AKSZ prescription \cite{AKSZ97} as developed by Cattaneo and Felder in \cite{CF01}. Firstly, a source manifold must be chosen, and this is taken to be the supermanifold $\Pi T\Sigma$, the tangent bundle of $\Sigma$ with reversed parity of the fiber. On this supermanifold we have a homological vector field $D$ of degree $1$ and a nondegenerate $D$-invariant measure $\mu$ of degree $-2$. Choosing local coordinates $\{ \sigma^{\alpha}\}$ on $\Sigma$ together with their odd counterparts $\{ \theta^{\alpha} \}$, we can write $D=\theta^{\alpha}\partial/\partial \sigma^{\alpha}$ and $\mu=\mathrm{d}^2\sigma \mathrm{d}^2 \theta$. The second input is the target, which is taken to be the supermanifold $\Pi T^*E^*$, the cotangent bundle of $E^*$ with reversed parity of the fiber. This is a graded symplectic manifold of degree $1$ with a canonical odd symplectic form $\omega$ that, using local coordinates $\{ x^{\mu},\lambda_{I} \}$ on $E^{*}$ and their odd counterparts $\{\eta_{\mu},\psi^{I}\}$, can be written as
$$
\omega=\d \eta_{\mu} \d x^{\mu}+\d \psi^{I} \d \lambda_{I}.
$$
We also have a homological vector field $Q$ of degree $1$. It is derived from the Poisson bivector field on $E^*$. A formula for it in the same coordinates is
\begin{align*}
Q= \rho^{\mu}_I\psi^I\frac{\partial}{\partial x^{\mu}}+\big( \rho^{\mu}_I\eta_{\mu}+C^K_{IJ}\lambda_K\psi^J\big)\frac{\partial}{\partial \lambda_I} +\left(\partial_{\mu} \rho^{\nu}_I\psi^I\eta_{\nu}+ \frac{1}{2}\partial_{\mu}C^{K}_{IJ}\lambda_K\psi^I\psi^K\right)\frac{\partial}{\partial \eta_{\mu}}+\frac{1}{2}C^I_{JK}\psi^J\psi^K\frac{\partial}{\partial \psi^I},
\end{align*}
where $\rho^{\mu}_I$ and $C^K_{IJ}$ are the structure functions of the Lie algebroid. One can check that $Q$ preserves $\omega$ and the corresponding Hamiltonian of degree $2$ is
$$
\Theta= \rho^{\mu}_I\psi^I\eta_{\mu}+\frac{1}{2}C^K_{IJ}\lambda_K\psi^I\psi^J.
$$ 

In the AKSZ formalism, a field configuration is defined by what is called a {\it superfield}, that is, a map $\Phi \colon \Pi T\Sigma \to \Pi T^*E^*$.\footnote{The phrase ``let $\Phi$ be a map from $\Pi T\Sigma$ to $\Pi T^*E^*$'' is an abuse of language for ``let $B$ be a supermanifold and let $\Phi$ be a map from $\Pi T\Sigma \times B$ to $\Pi T^*E^*$''. See section~2.3 of \cite{Roytenberg07} for a fuller explanation.} We can view $\Phi$ as a map $(\gbf{\phi},\gbf{\lambda})$ from $\Pi T\Sigma$ to $E^*$ together with a section $(\gbf{\eta},\gbf{\psi})$ of the odd vector bundle $(\gbf{\phi},\gbf{\lambda})^*\Pi T^*E^*$. We write $\Phi=(\gbf{\phi},\gbf{\lambda},\gbf{\eta},\gbf{\psi})$ with this understanding as to the meaning of $(\gbf{\phi},\gbf{\lambda})$ and $(\gbf{\eta},\gbf{\psi})$. Notice also that the space of superfields inherits a grading from $\Pi T^*E^*$, which is commonly referred to as {\it ghost number}. In terms of local coordinates on $ \Pi T\Sigma$,
\begin{align*}
\gbf{\phi}&=\phi+\theta^{\alpha}\eta^{+}_{\alpha}-\frac{1}{2}\theta^{\alpha}\theta^{\beta}\zeta^{+}_{\alpha\beta},\\
\gbf{\lambda}&=\lambda+\theta^{\alpha}\psi^{+}_{\alpha}-\frac{1}{2}\theta^{\alpha}\theta^{\beta}\xi^{+}_{\alpha\beta},\\
\gbf{\eta}&=\zeta+\theta^{\alpha}\eta_{\alpha}+\frac{1}{2}\theta^{\alpha}\theta^{\beta}\phi^{+}_{\alpha\beta},\\
\gbf{\psi}&=\xi+\theta^{\alpha}\psi_{\alpha}+\frac{1}{2}\theta^{\alpha}\theta^{\beta}\lambda^{+}_{\alpha\beta}.
\end{align*}
In the expansion, $\phi$ is a map $\Sigma \to X$, $\lambda$ is a section of $\phi^*E^*$, $\eta$ is a one-form on $\Sigma$ with values in $\phi^*T^*X$, and $\psi$ is a one-form on $\Sigma$ with values in $\phi^*E$. These fields comprise precisely the field content of the Lie algebroid Poisson sigma model of section~\ref{Sect:3} and they all have ghost number zero. The {\it ghost fields} are $\zeta$, a section of $\phi^*T^*X$, and $\xi$, a section of $\phi^*E$; they both have ghost number one. The other fields $\phi^+$, $\lambda^+$, $\eta^+$, $\psi^+$, $\zeta^+$, and $\xi^+$ are the {\it antifields} of $\phi$, $\lambda$, $\eta$, $\psi$, $\zeta$, and $\xi$, with ghost numbers $-1$, $-1$, $-1$, $-1$, $-2$, and $-2$, respectively.  

The space of superfields $\Phi=(\gbf{\phi},\gbf{\lambda},\gbf{\eta},\gbf{\psi})$ is endowed with an odd symplectic form $\Omega$ of ghost number $-1$, which is obtained from the symplectic form $\omega$ on $\Pi T^*E^*$ upon integration over $\Pi T\Sigma$. Writing things out in local coordinates on $\Pi T^*E^*$, we have
$$
\Omega=\int_{\Pi T\Sigma}\mathrm{d}^2\sigma \mathrm{d}^2 \theta \big(\mathbf{d} \gbf{\eta}_{\mu} \mathbf{d} \gbf{\phi}^{\mu}+\mathbf{d}\gbf{\psi}^{I} \mathbf{d} \gbf{\lambda}_{I}\big).
$$
Here the symbol $\mathbf{d}$ stands for the exterior derivative on the space of superfields. Just as in the usual case, one can define odd Poisson brackets $(\,,)$ acting on functionals of the superfields. These Poisson brackets, which are the Batalin-Vilkovisky antibrackets, obey a graded Jacobi identity. The space of superfields is also equipped with a homological vector field $\delta$ of ghost number $1$ preserving $\Omega$, which we call the {\it BRST operator}. It is defined as the sum of the commuting vector fields $\widehat{D}$ and $\widehat{Q}$ obtained from $D$ and $Q$ by acting on superfields $\Phi\colon \Pi T\Sigma \to \Pi T^*E^*$ by the corresponding infinitesimal diffeomorphisms of $\Pi T\Sigma$ on the left and of $\Pi T^*E^*$ on the right. The Hamiltonian function $S$ (of ghost number zero) of this vector field is then the Batalin-Vilkovisky action functional. More explicitly, this is
\begin{equation}
\label{eq:5.1}
S=\int_{\Pi T\Sigma}\mathrm{d}^2\sigma \mathrm{d}^2 \theta \left(\gbf{\eta}_{\mu} D \gbf{\phi}^{\mu}+\gbf{\psi}^I D \gbf{\lambda}_I + \rho^{\mu}_I(\gbf{\phi})\gbf{\psi}^I  \gbf{\eta}_{\mu}+\frac{1}{2}C^K_{IJ}(\gbf{\phi})\gbf{\lambda}_K \gbf{\psi}^I \gbf{\psi}^J\right).
\end{equation}
Performing the $\mathrm{d}^2 \theta$ integration, we find
\begin{align*}
S&=\int_{\Sigma}\eta_{\mu} \wedge \d \phi^{\mu}+\psi^I \wedge \d \lambda_I + \rho^{\mu}_I\psi^I \wedge \eta_{\mu}+\frac{1}{2}C^K_{IJ}\lambda_K \psi^I \wedge \psi^J \\
&\qquad +\zeta_{\mu} \d \eta^{+\mu}+\xi^I \d \psi^{+}_I+\psi^+_K \wedge \psi^I C^K_{IJ}\xi^J+\eta^{+\mu}\wedge \eta_{\mu}\partial_{\nu}\rho^{\mu}_I\xi^I+\eta^{+\mu}\wedge \psi^I \partial_{\nu}\rho^{\mu}_I\zeta_{\mu}\\
&\qquad +\eta^{+\mu}\wedge \psi^I \partial_{\mu}C^K_{IJ}\lambda_K\xi^J+\frac{1}{2}\eta^{+\mu}\wedge \psi^+_K  \partial_{\mu}C^K_{IJ} \xi^I\xi^J-\frac{1}{2}\eta^{+\rho}\wedge \eta^{+\nu} \partial_{\rho} \partial_{\nu}\rho^{\mu}_I \xi^I\zeta_{\mu}\\
&\qquad -\frac{1}{2}\zeta^{+\nu}\partial_{\nu}\rho^{\mu}_I \xi^I\zeta_{\mu}+\frac{1}{2}\phi^{+}_{\mu}\rho^{\mu}_I\xi^I+\frac{1}{2}\lambda^{+I}\rho^{\mu}_I\zeta_{\mu}-\frac{1}{4}\zeta^{+\nu}\partial_{\nu}C^K_{IJ}\lambda_K\xi^I\xi^J-\frac{1}{4}\xi^+_K C^K_{IJ}\xi^I\xi^J \\
&\qquad -\frac{1}{4}\eta^{+\mu}\wedge\eta^{+\nu}  \partial_{\mu} \partial_{\nu}C^K_{IJ}\lambda_K\xi^I\xi^J.
\end{align*}
Comparing this with the expression for the classical action $S_0$ in \eqref{eq:3.1}, we see that the latter is recovered by setting the antifields in $S$ to zero. 

If $\Sigma$ has no boundary, the condition that the BRST operator $\delta$ be nilpotent is equivalent to what is called the {\it classical master equation}:
$$
(S,S)=0.
$$
Therefore the Batalin-Vilkovisky action $S$ is automatically BRST-invariant, with the BRST transformation rules
\begin{align*}
\begin{split}
\delta \gbf{\phi}^{\mu}&=D \gbf{\phi}^{\mu}+\rho^{\mu}_I( \gbf{\phi}) \gbf{\psi}^{I},\\
\delta  \gbf{\lambda}_{I}&=D \gbf{\lambda}_{I}+\rho^{\mu}_I( \gbf{\phi})\gbf{\eta}_{\nu}+C^K_{IJ}(\gbf{\phi})\gbf{\lambda}_K\gbf{\psi}^J,\\
\delta \gbf{\eta}_{\mu} &=D\gbf{\eta}_{\mu} +\partial_{\mu}\rho^{\nu}_I( \gbf{\phi})\gbf{\psi}^{I} \gbf{\eta}_{\nu}+\frac{1}{2}\partial_{\mu}C^I_{JK}(\gbf{\phi})\gbf{\lambda}_{I}\gbf{\psi}^J\gbf{\psi}^K,\\
\delta \gbf{\psi}^I &=D\gbf{\psi}^I+\frac{1}{2}C^I_{JK}(\gbf{\phi})\gbf{\psi}^J\gbf{\psi}^K.
\end{split}
\end{align*}
To evaluate the effect of the BRST operator on general functionals, we note that the antibracket acts as a derivation, in the sense that
$$
(F,GH)=(F,G)H \pm G(F,H),
$$
where the sign is $-1$ if the parity of the ghost number of $G$ is odd and that of $F$ is even, and $+1$ otherwise. Hence if $G$ and $H$ are arbitrary functionals, then
$$
\delta(GH)=\delta G H \pm G \delta H,
$$ 
where the sign is $+$ or $-$ if the parity of the ghost number of $G$ is even or odd. 

To close this longish section, we discuss the case when $\Sigma$ has a boundary, which we assume to consist of a single circle for simplicity. In such a case, appropriate boundary conditions must be imposed in order that the classical master equation $(S,S)=0$ be identically satisfied. According to section~3.2 of \cite{CQZ10}, the boundary conditions are that $\Phi \colon \Pi T\Sigma \to \Pi T^*E^*$ restricts on the boundary to a map $\Pi T\partial \Sigma \to \Pi N^* F^{\perp}$ for a Lie subalgebroid $F$ of $E$ over a submanifold $Y \subset X$. If one fixes adapted local coordinates $\{ x^{\mu'},x^{\mu''} \}$ and a local trivialization $\{ e_{I'},e_{I''}\}$ as in the preceding section, these boundary conditions can be rephrased by saying that $\gbf{\phi}^{\mu''}=0$, $\gbf{\lambda}_{I'}=0$, $\gbf{\eta}_{\mu'}=0$, and $\gbf{\psi}^{I''}=0$ on $\Pi T\partial\Sigma$. This can be made more explicit by using local coordinates $\{ \tau,\theta \}$ on $\Pi T\partial \Sigma$ and writing $\gbf{\phi}^{\mu}$, $\gbf{\lambda}_{I}$, $\gbf{\eta}_{\mu}$, and $\gbf{\psi}^{I}$ in terms of components:
\begin{alignat*}{2}
\gbf{\phi}^{\mu}&=\phi^{\mu}+\theta\eta^{+\mu}_{\tau}, & \qquad \gbf{\lambda}_{I}&=\lambda_I+\theta \psi^{+}_{I\tau},\\
\gbf{\eta}_{\mu}&=\zeta_{\mu}+\theta\eta_{\mu\tau}, & \qquad \gbf{\psi}^I&=\xi^I+\theta\psi^{I}_{\tau}.
\end{alignat*}
(The indices $\mu$ and $I$ here collectively denote the primed and double-primed indices.) The above boundary conditions are then translated to
\begin{equation*}
\phi^{\mu''}=0,\qquad\lambda_{I'}=0,\qquad\eta_{\mu' \tau}=0,\qquad\psi^{I''}_{\tau}=0,\qquad\zeta_{\mu'}=0,\qquad \xi^{I''}=0, \qquad \eta^{+\mu''}_{\tau}=0, \qquad  \psi^{+}_{I'\tau}=0.
\end{equation*}
Excluding the antifields, these are precisely the boundary conditions that we found in section~\ref{Sect:BC}.

\section{Boundary observables}
\label{Sect:6}
In this section, we will discuss the boundary observables which can be defined in the Lie algebroid Poisson sigma model described in section~\ref{Sect:AKSZ}. We will restrict ourselves to observables in the BRST sense.  

On general grounds, the classical observables for a Lie algebroid Poisson sigma model are classified according to the cohomology of the BRST operator $\delta$. As explained in detail in \cite{CQZ10}, the latter is identified with the cohomology of the odd vector field $Q$ acting on the algebra of functions on $\Pi T^*E^*$. By using the results of \cite{BZ09} one can therefore establish that the classical observables correspond exactly to the Lichnerowicz-Poisson cohomology of $E^*$. 

Consider now a boundary condition corresponding to a Lie subalgebroid $F$ over a submanifold $Y \subset X$. The above suggests that the natural observables in this context are related to certain cohomology classes on $F$. To see how this comes about, it is convenient to make the following general remarks. 

According to the Theorem~7.2 of \cite{Vor01}, there is a diffeomorphism $\Pi T^*E^* \to \Pi T^*(\Pi E)$ preserving the odd symplectic structure, which in local coordinates is given by $(x^{\mu},\lambda_I,\eta_{\mu},\psi^I)\mapsto (x^{\mu},-\psi^{I},\eta_{\mu},-\lambda_{I})$. Under this diffeomorphism, the Hamiltonian $\Theta$ becomes
$$
\Theta= -\rho^{\mu}_I\psi^I\eta_{\mu}+\frac{1}{2}C^K_{IJ}\lambda_K\psi^I\psi^J,
$$ 
while the restriction of $Q$ to $\Pi E$ reads
$$
Q_E=-\rho^{\mu}_{I}\psi^{I}\frac{\partial}{\partial x^{\mu}}+\frac{1}{2}C^{K}_{IJ}\psi^{I}\psi^{J}\frac{\partial}{\partial \psi^{K}}.
$$
Like $Q$, the odd vector field $Q_E$ is nilpotent, so it defines a coboundary operator on the algebra of functions on $\Pi E$. The cohomology associated with $Q_E$ can easily be shown to be isomorphic to the Lie algebroid cohomology of $E$; see \cite{Vaintrob97} for details (beware that the expression for $Q_E$ differ by a minus sign from that of \cite{Vaintrob97}). In the same vein, one can further restrict $Q_E$ to $\Pi F$ to obtain a homological vector field $Q_F$ on $\Pi F$ whose cohomology yields the Lie algebroid cohomology of $F$.

From now on we identify $\Pi T^* E^*$ with $\Pi T^*(\Pi E)$ by means the above diffeomorphism. With this identification, the boundary conditions require that $\Phi=(\gbf{\phi},\gbf{\psi},\gbf{\eta},\gbf{\lambda})$ maps $\Pi T\partial \Sigma$ to $\Pi N^*(\Pi F)$. In the local coordinates and trivialization of section~\ref{Sect:BC}, this means that $\gbf{\eta}_{\mu'}$, and $\gbf{\lambda}_{I'}$ are subject to the same conditions as before. In other words, we have $\gbf{\eta}_{\mu'}=0$ and $\gbf{\lambda}_{I'}=0$ on $\Pi T\partial\Sigma$. This makes it clear that nontrivial boundary observables must come from functions on the base $\Pi F$ of $\Pi N^*(\Pi F)$. The residual BRST transformation rules are the following:
\begin{align}
\label{eq:6.1}
\begin{split}
\delta \gbf{\phi}^{\mu'}&=D \gbf{\phi}^{\mu'}-\rho^{\mu'}_{I'}( \gbf{\phi}^{\mu'},0) \gbf{\psi}^{I'},\\
\delta \gbf{\psi}^{I'} &=D\gbf{\psi}^{I'}+\frac{1}{2}C^{I'}_{J'K'}( \gbf{\phi}^{\mu'},0)\gbf{\psi}^{J'}\gbf{\psi}^{K'},
\end{split}
\end{align}
where $D$ here denotes the odd vector field $\theta \partial_{\tau}$ on $\Pi T\partial\Sigma$. For later use, we also record the residual BRST transformation rules for the component fields:
\begin{align}
\label{eq:bdyBRST}
\begin{split}
\delta \phi^{\mu'}&=-\rho^{\mu'}_{I'}( \phi^{\mu'},0)\xi^{I'},\\
\delta \eta^{+\mu'}_{\tau}&=-\partial_{\tau}\phi^{\mu'}+\rho^{\mu'}_{I'}( \phi^{\mu'},0)\psi^{I'}_{\tau}+\partial_{\nu'}\rho^{\mu'}_{I'}( \phi^{\mu'},0)\eta^{+\nu'}_{\tau},\\
\delta \xi^{I'}&=\frac{1}{2}C^{I'}_{J'K'}(\phi^{\mu'},0)\xi^{J'}\xi^{K'},\\
\delta \psi^{I'}_{\tau}&=-\partial_{\tau}\xi^{I'}-C^{I'}_{J'K'}(\phi^{\mu'},0)\psi^{J'}_{\tau}\xi^{K'}-\frac{1}{2}\partial_{\nu'}C^{I'}_{J'K'}(\phi^{\mu'},0)\eta^{+\nu'}_{\tau}\xi^{J'}\xi^{K'}.
\end{split}
\end{align}

We now wish to construct the boundary observables which are invariant under the BRST transformations \eqref{eq:6.1}. From our previous discussion, we know that the interesting boundary observa\-bles are induced by functions on $\Pi F$. A function on $\Pi F$ of degree $k$ is of the general form $\omega=\frac{1}{k!}\omega_{I'_1 \cdots I'_k}e^{I'_1}\wedge \cdots \wedge e^{I'_k}$ and corresponds to an $F$-form of degree $k$. For every such $F$-$k$-form $\omega$, we consider the functional
$$
 \mathcal{O}_{\omega}=\frac{1}{k!}\omega_{I'_1 \cdots I'_k}(\gbf{\phi})\gbf{\psi}^{I'_1}\cdots\gbf{\psi}^{I'_k}.
$$
Written out explicitly, this is 
$$
 \mathcal{O}_{\omega}= \mathcal{O}_{\omega}^{(0)}+\theta \mathcal{O}_{\omega\tau}^{(1)},
$$
where
\begin{align}
\label{eq:6.3}
\begin{split}
 \mathcal{O}_{\omega}^{(0)}&= \frac{1}{k!}\omega_{I'_1 \cdots I'_k}\xi^{I'_1}\cdots \xi^{I'_k},\\
 \mathcal{O}_{\omega\tau}^{(1)}&=\frac{1}{(k-1)!}\omega_{I'_1 \cdots I'_k}\xi^{I'_1}\cdots \xi^{I'_{k-1}}\psi^{I'_k}_{\tau}+\frac{1}{k!}\eta^{+\nu'}_{\tau}\partial_{\nu'}\omega_{I'_1 \cdots I'_k}\xi^{I'_1}\cdots \xi^{I'_k}.
\end{split}
\end{align}
We need to calculate the change of $ \mathcal{O}_{\omega}$ under a BRST transformation. It is a simple matter to see that
$$
\delta \mathcal{O}_{\omega}=D \mathcal{O}_{\omega}+ \mathcal{O}_{Q_F \omega}.
$$
In components this becomes
\begin{align}
\label{eq:6.4}
\begin{split}
\delta \mathcal{O}_{\omega}^{(0)}&= \mathcal{O}_{Q_F \omega}^{(0)},\\
\delta \mathcal{O}_{\omega}^{(1)}&=-\d \mathcal{O}_{\omega}^{(0)}- \mathcal{O}_{Q_F \omega}^{(1)}.
\end{split}
\end{align}
From the first equation in \eqref{eq:6.4} we learn that $\mathcal{O}_{\omega}^{(0)}$ is BRST-invariant if and only if $\omega$ is a $Q_F$-closed $F$-$n$-form. Moreover, if $\omega$ is a $Q_F$-exact $F$-$n$-form, then the corresponding functional $\mathcal{O}_{\omega}^{(0)}$ is $\delta$-exact. Thus, the BRST cohomology classes of functionals obtainable this way correspond exactly to the Lie algebroid cohomology classes on $F$.  In addition, note that if we take $\omega$ to be a representative of a Lie algebroid cohomology class, the second equation in \eqref{eq:6.4} reduces to
\begin{equation}
\label{eq:6.5}
\delta \mathcal{O}_{\omega}^{(1)}=-\d \mathcal{O}_{\omega}^{(0)}.
\end{equation}
Together with $\delta \mathcal{O}_{\omega}^{(0)}=0$, this is equivalent to what is known in the literature as a {\it descent equation}.

In light of the foregoing remarks, the most obvious BRST-invariant boundary observable is
$$
 \mathcal{O}_{\omega}^{(0)}(P)= \frac{1}{k!}\omega_{I'_1 \cdots I'_k}(\phi^{\mu'}(P))\xi^{I'_1}(P)\cdots \xi^{I'_k}(P),
$$
where $P$ is an arbitrary point in $\partial \Sigma$. It is easy to verify that equation \eqref{eq:6.5} implies that this is independent of the choice of $P$. In a similar manner $ \mathcal{O}_{\omega}^{(1)}$ can be used to obtain a boundary observable, by
\begin{align*}
\int_{\partial \Sigma} \mathcal{O}_{\omega}^{(1)}&=\int_{\partial \Sigma}\left\{\frac{1}{(k-1)!}\omega_{I'_1 \cdots I'_k}(\phi^{\mu'}(\tau))\xi^{I'_1}(\tau)\cdots \xi^{I'_{k-1}}(\tau)\psi^{I'_k}_{\tau} \right.\\
&\qquad\qquad\left.+\frac{1}{k!}\eta^{+\nu'}_{\tau}\partial_{\nu'}\omega_{I'_1 \cdots I'_k}(\phi^{\mu'}(\tau))\xi^{I'_1}(\tau)\cdots \xi^{I'_k}(\tau)\right\} \d \tau.
\end{align*}
Since according to \eqref{eq:6.5}, $ \mathcal{O}_{\omega}^{(1)}$ is BRST-invariant up to a total derivative, this expression is BRST-invariant. It is also worthwhile to mention that this may be used to deform the Batalin-Vilkovisky action (at least to first order):
$$
S \to S+\epsilon\int_{\partial \Sigma} \mathcal{O}_{\omega}^{(1)},
$$
where $\epsilon$ is a formal parameter. The BRST-invariant boundary observables are then given by arbitrary products of the $ \mathcal{O}_{\omega}^{(0)}(P)$ and the integrated observable $\int_{\partial \Sigma} \mathcal{O}_{\omega}^{(1)}$.

\section{Coupling to $F$-connections}
\label{Sect:7}
So far, we have seen how to describe boundary conditions in Lie algebroid Poisson sigma models as Lie subalgebroids $F$ of the target algebroid $E$ supported on a submanifold $Y \subset X$. In this section, we will consider the coupling of these models to ordinary $F$-connections on a vector bundle over $Y$. We wish to carry out this coupling in a way that preserves the BRST symmetry \eqref{eq:bdyBRST}. For simplicity of exposition, we suppose that the boundary of $\Sigma$ is a single circle parametrized by $0 \leq \tau \leq 1$ (with $\tau=0$ and $\tau=1$ identified).

Let $\nabla$ be an $F$-connection on a vector bundle $V$ over $Y$. In the trivialization and local coordinates introduced in section~\ref{Sect:BC}, such a connection is concretely given by specifying $\End(V)$-valued functions $A_I(x^{\mu})$.\footnote{Here, in order to lighten the notation, we have omitted the prime on the indices $\mu$ and $I$.} Using these, we can define a connection one-form on the bundle $\phi^*V$ over $\partial \Sigma$ by means of
\begin{equation*}
M(\tau)=A_I(\phi^{\mu}(\tau)) \psi^I_{\tau}+\eta^{+\nu}_{\tau}\partial_{\nu}A_I(\phi^{\mu}(\tau))\xi^I,
\end{equation*}
where $\eta^{+\nu}_{\tau}$ is the antifield we encountered in sections \ref{Sect:AKSZ} and \ref{Sect:6}. Consider the parallel transport opera\-tor of this connection along the path $0 \leq \tau \leq t$ in the boundary $\partial \Sigma$, which can be written as a path-ordered exponential
\begin{align*}
U(0,t) &= P\!\exp \left( -\int_0^t M(\tau)\d \tau\right) \\
&=1 +\sum_{n=1}^{\infty}(-1)^n \int_0^t M(\tau_1) \int_0^{\tau_1} M(\tau_2) \cdots \int_0^{\tau_{n-1}}M(\tau_n) \d \tau_n \cdots \d \tau_2 \d \tau_1.
\end{align*}
It satisfies the first order differential equation 
\begin{equation}
\label{eq:7.1}
\partial_{\tau}U(0,\tau)=-M(\tau)U(0,\tau)
\end{equation}
and the initial condition $U(0,0)=1$. The final value $U(0,1)$ is regarded as the holonomy operator around $\partial \Sigma$.

To couple the Lie algebroid Poisson sigma model of section~\ref{Sect:AKSZ} to the $F$-connection $\nabla$, we shall follow the approach put forward by Witten in \cite{Witt95}. First of all, let us note that the quantum theory of this model is formally given by a Feynman path integral  
\begin{equation}
\label{eq:7.2}
\int_{\mathfrak{L}} D\Phi \exp(-S),
\end{equation}
where $\mathfrak{L}$ denotes a Lagrangian submanifold in the space of superfields $\Phi\colon \Pi T\Sigma \to \Pi T^*E^*$ subject to the boundary condition that $\Pi T\partial \Sigma$ is mapped to $\Pi N^*F^{\perp}$.\footnote{The choice of $\mathfrak{L}$ is called a gauge fixing and it is typically generated by a gauge fermion $\Psi$; see \cite{AKSZ97,Hen92,Sch93,Witt90} for background.} With this understanding, the coupling to the $F$-connection $\nabla$ is obtained by modifying the path integral to
\begin{equation}
\label{eq:7.3}
\int_{\mathfrak{L}} D\Phi \exp(-S) \cdot \tr U(0,1)=\int_{\mathfrak{L}} D\Phi \exp(-S)\cdot \tr P\!\exp \left( -\int_0^1 M(\tau)\d \tau\right),
\end{equation}
with `$\tr$' here denoting the trace on $\End(V_{\phi(0)})$. The factor $\tr U(0,1)$ can be thought of as the addition of a boundary interaction; see for instance \cite{KrLar01,TTU01}.

We now want to determine the condition on $\nabla$ that ensures invariance under the BRST transformations \eqref{eq:bdyBRST}. This will be done using the methods of \cite{Laz05}. We start by noting that, by a straightforward calculation, the change under a BRST transformation in the trace of the holonomy about $\partial \Sigma$ is
 \begin{equation}
 \label{eq:7.4}
\delta \tr U(0,1)=-\tr \left( U(0,1)\int_0^1 U(0,\tau)^{-1}\delta M(\tau) U(0,\tau)\d \tau\right).
\end{equation}
In order to have BRST invariance, we must set $\delta \tr U(0,1)$ equal to zero.

Our task is now reduced to calculating the change in $M$ produced by a BRST transformation. Using the general rules outlined in section~\ref{Sect:AKSZ}, we find
\begin{align}
\label{eq:7.5}
\begin{split}
\delta M &=-\partial_{\tau}\big(A_I \xi^I\big) -\big(\rho^{\mu}_I\partial_{\mu}A_J-\rho^{\mu}_J \partial_{\mu}A_I-C^K_{IJ}A_K\big)\xi^I\psi_{\tau}^J\\
&\quad +\frac{1}{2}\eta^{+\nu}_{\tau}\partial_{\nu}\big(\rho^{\mu}_I\partial_{\mu}A_J-\rho^{\mu}_J \partial_{\mu}A_I-C^K_{IJ}A_K\big)\xi^I\xi^J.
\end{split}
\end{align}
The next step is to observe that
\begin{align}
\label{eq:7.6}
\begin{split}
U(0,\tau)^{-1}\partial_{\tau}\big(A_I\xi^I\big)U(0,\tau) =\partial_{\tau}\big(U(0,\tau)^{-1}A_I\xi^IU(0,\tau)\big)+U(0,\tau)^{-1}\big[A_I\xi^I,M\big]U(0,\tau).
\end{split}
\end{align}
This follows readily from \eqref{eq:7.1}. In equation \eqref{eq:7.6}, the bracket $[,]$ denotes the usual commutator. One can easily verify that
\begin{equation}
\label{eq:7.7}
\big[A_I\xi^I,M\big]=[A_I,A_J]\xi^I \psi^J_{\tau}-\frac{1}{2}\eta^{+\nu}_{\tau}\partial_{\nu}[A_I,A_J]\xi^I\xi^J.
\end{equation}
Putting together equations \eqref{eq:7.5}, \eqref{eq:7.6} and \eqref{eq:7.7}, we have
\begin{align}
\label{eq:7.8}
\begin{split}
U(0,\tau)^{-1}\delta M(\tau) U(0,\tau) &= -\partial_{\tau}\big(U(0,\tau)^{-1}A_I(\phi^{\mu}(\tau))\xi^I(\tau)U(0,\tau)\big) \\
&\quad+U(0,\tau)^{-1}\left(-F_{IJ}(\phi^{\mu}(\tau))\xi^I\psi_{\tau}^J+\frac{1}{2}\eta^{+\nu}_{\tau}\partial_{\nu}F_{IJ}(\phi^{\mu}(\tau))\xi^I\xi^J \right)U(0,\tau),
\end{split}
\end{align}
where $F_{IJ}(\phi^{\mu}(\tau))$ are the components of the pullback by $\phi$ of the $F$-curvature of $\nabla$. We now substitute this expression back into \eqref{eq:7.4} to get
\begin{align}
\label{eq:7.9}
\begin{split}
\delta \tr U(0,1) = -\tr\left\{U(0,1)\int_0^1 U(0,\tau)^{-1}\left(-F_{IJ}(\phi^{\mu}(\tau))\xi^I\psi_{\tau}^J  +\frac{1}{2}\eta^{+\nu}_{\tau}\partial_{\nu}F_{IJ}(\phi^{\mu}(\tau))\xi^I\xi^J \right)U(0,\tau)\d \tau\right\}.
\end{split}
\end{align}
Here we have used the periodicity of $\phi^{\mu}$ and $\xi^I$ to conclude that the contribution of the total derivative term in \eqref{eq:7.8} is zero: 
\begin{equation*}
\tr \big(A_I(\phi^{\mu}(1))\xi^I(1) U(0,1)\big)-\tr\big(U(0,1)A_I(\phi^{\mu}(0))\xi^I(0)\big)=0.
\end{equation*}
From \eqref{eq:7.9}, we see immediately that the condition that $\tr U(0,1)$ be BRST-invariant requires that $F_{IJ}=0$, or, in other words, that the $F$-connection $\nabla$ should be a representation of $F$ on the vector bundle $V$.

\section{Boundary observables coupled to $F$-connections}
\label{Sect:8}
We will now like to describe the BRST-invariant boundary observables that can be defined after coupling to an $F$-connection. All notational conventions introduced in the previous section remain in force. 

In analogy with the construction of section~\ref{Sect:6}, we first associate a functional $\mathcal{O}_{\omega}$ to each $\End(V)$-valued $F$-$k$-form $\omega=\frac{1}{k!}\omega_{I_1 \cdots I_k}e^{I_1}\wedge \cdots \wedge e^{I_k}$, defined by
$$
 \mathcal{O}_{\omega}=\frac{1}{k!}\omega_{I_1 \cdots I_k}(\gbf{\phi})\gbf{\psi}^{I_1}\cdots\gbf{\psi}^{I_k}.
$$
In terms of components this is 
$$
\mathcal{O}_{\omega}= \mathcal{O}_{\omega}^{(0)}+\theta  \mathcal{O}_{\omega \tau}^{(1)},
$$
with $\mathcal{O}_{\omega}^{(0)}$ and $\mathcal{O}_{\omega\tau}^{(1)}$ being given by the same expressions as in \eqref{eq:6.3}. It thus follows trivially that the change on these components induced by a BRST transformation is the same as \eqref{eq:6.4}. Let us also remark that the ghost number of $\mathcal{O}_{\omega}^{(0)}$ is $k$, while that of $\mathcal{O}_{\omega \tau}^{(1)}$ is $k-1$. 

We must now examine how the presence of the $F$-connection $\nabla$ modifies the BRST transformation laws of \eqref{eq:6.4}. Here we may borrow from the analysis of \cite{HerbLaz05}. To start off, notice that (just as in section~\ref{Sect:6}) functionals of the form $\mathcal{O}_{\omega}^{(0)}$ can be inserted at any point in the boundary of $\Sigma$. Bearing this in mind, we generalize \eqref{eq:7.3} to consider 
\begin{equation}
\label{eq:8.1}
\int_{\mathfrak{L}} D\Phi \exp(-S) \cdot \str \left(H(t)\mathcal{O}_{\omega}^{(0)}(t)\right).
\end{equation}
Here $H(t)$ denotes the conjugacy class of the holonomy $U(0,1)$ around the boundary $\partial \Sigma$,
\begin{equation*}
H(t)=U(0,t)U(0,1)U(0,t)^{-1},
\end{equation*}
and `$\str$' denotes the supertrace on $\End(V_{\phi(t)})\otimes \Lambda^{\bullet} F^*_{\phi(t)}$. Let us determine the condition on $\mathcal{O}_{\omega}^{(0)}$ for \eqref{eq:8.1} to be BRST-invariant.  

We have seen in the last section that BRST invariance of the boundary coupling \eqref{eq:7.3} is restored if we demand that the $F$-connection $\nabla$ is flat. In this case, \eqref{eq:7.8} becomes simply
\begin{equation}
\label{eq:8.2}
U(0,\tau)^{-1}\delta M(\tau) U(0,\tau) = -\partial_{\tau}\big(U(0,\tau)^{-1}A_I(\phi^{\mu}(\tau))\xi^I(\tau)U(0,\tau)\big).
\end{equation}
On the other hand, a BRST transformation will produce in $H(t)$ a change
\begin{equation}
\label{eq:8.3}
\delta H(t)=-U(0,t+1)\left(\int_t^{t+1} U(0,\tau)^{-1}\delta M(\tau) U(0,\tau) \d \tau\right) U(0,t)^{-1}.
\end{equation}
Using \eqref{eq:8.2} in \eqref{eq:8.3} gives
\begin{equation}
\label{eq:8.4}
\delta H(t)=\big[ A_I(\phi^{\mu}(t))\xi^{I}(t),H(t)\big].
\end{equation}
Taking note of \eqref{eq:8.4}, one easily obtains the variational formula
\begin{equation}
\label{eq:8.5}
\delta \str \left(H(t)\mathcal{O}_{\omega}^{(0)}(t)\right)=\str\left( H(t)\widehat{\delta}\mathcal{O}_{\omega}^{(0)}(t)\right),
\end{equation}
where we have set
\begin{equation*}
\widehat{\delta}\mathcal{O}_{\omega}^{(0)}=\delta\mathcal{O}_{\omega}^{(0)}-\xi^I\big[ A_I, \mathcal{O}_{\omega}^{(0)} \big].
\end{equation*}
It follows then immediately from \eqref{eq:8.5} that the invariance of \eqref{eq:8.1} under a BRST transformation requires that
\begin{equation}
\label{eq:8.6}
\widehat{\delta}\mathcal{O}_{\omega}^{(0)}=0.
\end{equation}

The crucial point is now that the operator $\widehat{\delta}$ is nilpotent. Indeed, a simple calculation shows that
$$
\widehat{\delta}^2 \mathcal{O}_{\omega}^{(0)}=\frac{1}{2}\big[ F_{IJ}\xi^I\xi^J,\mathcal{O}_{\omega}^{(0)} \big],
$$
which vanishes since $F_{IJ}=0$. Thus $\widehat{\delta}$ can be regarded as a ``covariant'' BRST operator with respect to the background flat $F$-connection $\nabla$. In particular, \eqref{eq:8.6} says that discussing the boundary observa\-bles in the present context requires us to consider ``covariant'' BRST cohomology classes. 

It is equally straightforward to verify the following analogs of \eqref{eq:6.4}:
\begin{align}
\label{eq:8.7}
\begin{split}
\widehat{\delta} \mathcal{O}_{\omega}^{(0)}&= \mathcal{O}_{Q_{F,\nabla} \omega}^{(0)},\\
\widehat{\delta} \mathcal{O}_{\omega}^{(1)}&=-\d \mathcal{O}_{\omega}^{(0)}- \mathcal{O}_{Q_{F,\nabla} \omega}^{(1)}.
\end{split}
\end{align}
Here $Q_{F,\nabla}$ is the homological vector field on $\Pi F$ defined by the same formula as $Q_F$ but with $\rho^{\mu}_I \partial_{\mu}$ replaced by the covariant derivative $\nabla_I=\rho^{\mu}_I\partial_{\mu} +[A_I,\,\,]$. The first equation in \eqref{eq:8.7} tells us that the functional $\mathcal{O}_{\omega}^{(0)}$ is $\widehat{\delta}$-closed if and only if $\omega$ is $Q_{F,\nabla}$-closed, and likewise $\mathcal{O}_{\omega}^{(0)}$ is $\widehat{\delta}$-exact if and only if $\omega$ is $Q_{F,\nabla}$-exact. Therefore, the boundary observables that can be constructed this way correspond exactly to the $Q_{F,\nabla}$-cohomology.   
As discussed for instance in \cite{Zuc08}, this is the Lie algebroid cohomology of $F$ with coefficients in $\End(V)$. We note further that when $\omega$ is taken to be a representative of a Lie algebroid cohomology class, the second equation in \eqref{eq:8.7} simply reads
\begin{equation}
\label{eq:8.8}
\widehat{\delta}\mathcal{O}_{\omega}^{(1)}=-\d \mathcal{O}_{\omega}^{(0)}.
\end{equation}
In other words, the descent equations are obtained by changing $\delta$ into $\widehat{\delta}$. From \eqref{eq:8.8}, it follows at once that
\begin{equation*}
\int_{\partial \Sigma}\mathcal{O}_{\omega}^{(1)},
\end{equation*}
is a $\widehat{\delta}$-invariant boundary observable. Just as in our earlier treatment in section~\ref{Sect:6}, this observable may be interpreted as a deformation of the boundary interaction. To be precise, one can modify the path integral \eqref{eq:7.2} by inserting
\begin{equation*}
\tr P\!\exp\left(\epsilon \int_{\partial \Sigma}\mathcal{O}_{\omega}^{(1)}\right),
\end{equation*}
where $\epsilon$ is regarded as a formal parameter.

\section{Coupling to $F$-superconnections}
\label{Sect:9}
In section~\ref{Sect:7}, we extended the AKSZ construction of section~\ref{Sect:AKSZ} to include coupling to $F$-connections on a vector bundle over $Y$. In the present section, we generalize this to the graded context; we will couple the Lie algebroid Poisson sigma model of section~\ref{Sect:AKSZ} to $F$-superconnections on a $\ZZ$-graded vector bundle over $Y$ in a BRST-invariant fashion. 

So let $V=\bigoplus_{k \in \ZZ}V^k$ be a $\ZZ$-graded vector bundle on $Y$ equipped with an $F$-superconnection $D$. As we saw in section~\ref{Sect:2}, the latter is determined by giving an $F$-connection $\nabla$ on $V$ which preserves the $\ZZ$-grading, and a $\Hom(V^{\bullet},V^{\bullet+1-k})$-valued $F$-$k$-form $\Omega^{(k)}$ for each positive integer $k \neq 1$. In the adapted local coordinates and trivialization discussed in section~\ref{Sect:BC}, $\nabla$ can be described via $\End^0(V)$-valued functions $A_I(x^{\mu})$, and $\Omega^{(k)}$ is expressed in terms of its $\Hom(V^{\bullet},V^{\bullet+1-k})$-valued components $\Omega^{(k)}_{I_1 \cdots I_k}(x^{\mu})$.\footnote{Here again, we shall lighten the notation by dropping the prime on the indices.} Keeping the notation of section~\ref{Sect:2}, we set $v = \Omega^{(0)}$.

Now let us consider a superfield $\Phi\colon \Pi T\Sigma \to \Pi T^*E^*$ with the property that it restricts on the boundary to a map $\Pi T\partial \Sigma \to \Pi N^*F^{\perp}$. We would like to include a boundary interaction as in \eqref{eq:7.2}. For this, we first note that there is an obvious map $\pi\colon \Pi N^*F^{\perp} \to Y$, and we may regard $v(x^{\mu})$, $A_I(x^{\mu})$ and $\Omega^{(k)}_{I_1 \cdots I_k}(x^{\mu})$ as locally-defined sections of the $\ZZ$-graded vector bundle $\pi^*\End(V)$. Hence, if we let $j$ denote the inclusion $\partial \Sigma \to \Pi T \partial \Sigma$, we may define a connection one-form on the bundle $(\Phi\circ j)^*V$ over $\partial \Sigma$ by setting
\begin{align*}
M(\tau) &=\eta^{+\nu}_{\tau}\partial_{\nu}v(\phi^{\mu}(\tau))+A_I(\phi^{\mu}(\tau)) \psi^I_{\tau}+\eta^{+\nu}_{\tau}\partial_{\nu}A_I(\phi^{\mu}(\tau))\xi^I \\
&\quad+ \sum_{k \geq 2} \left\{\frac{1}{(k-1)!}\Omega^{(k)}_{I_1\cdots I_k}(\phi^{\mu}(\tau))\xi^{I_1}\cdots \xi^{I_{k-1}}\psi^{I_k}_{\tau}+\frac{1}{k!}\eta^{+\nu}_{\tau}\partial_{\nu}\Omega^{(k)}_{I_1\cdots I_k}(\phi^{\mu}(\tau))\xi^{I_1}\cdots \xi^{I_{k}}\right\}.
\end{align*}
It is easy to show that the connection so defined is independent of the adapted local coordinates and trivialization employed. Notice that $v(x^{\mu})$, $A_I(x^{\mu})$ and $\Omega^{(k)}_{I_1 \cdots I_k}(x^{\mu})$ are considered to have ghost number $1$, $0$, and $1-k$, respectively, so that $M$ is of ghost number $0$. Let $U(0,t)$ describe the parallel transport operator, as before, and define the corresponding holonomy around the boundary $U(0,1)$. Following the logic of section~\ref{Sect:7}, the coupling to the $F$-superconnection $D$ is accomplished by replacing \eqref{eq:7.2} by 
\begin{equation}
\label{eq:9.1}
\int_{\mathfrak{L}} D\Phi \exp(-S) \cdot \str U(0,1)=\int_{\mathfrak{L}} D\Phi \exp(-S)\cdot \str P\!\exp \left( -\int_0^1 M(\tau)\d \tau\right),
\end{equation}
where `$\str$' denotes the supertrace on $\End(V_{\Phi(j(0))})$. Here again, we must think of $\str U(0,1)$ as the insertion of a boundary interaction.

At this point, we may proceed as in section~\ref{Sect:7}. First we calculate the change in $\str U(0,1)$ under a BRST transformation. This is
\begin{equation}
 \label{eq:9.2}
\delta \str U(0,1)=-\str \left( U(0,1)\int_0^1 U(0,\tau)^{-1}\delta M(\tau) U(0,\tau)\d \tau\right).
\end{equation}
Next, we wish to calculate the effect of a BRST transformation on $M$. This is a bit long, but straightforward. The result is
\begin{align}
\label{eq:9.3}
\begin{split}
\delta M &=-\partial_{\tau}\left(v+A_I \xi^I+\sum_{k\geq 2}\frac{1}{k!}\Omega^{(k)}_{I_1\cdots I_k}\xi^{I_1}\cdots\xi^{I_k} \right) \\
&\quad +\rho^{\mu}_I\partial_{\mu}v \psi^I_{\tau}-\eta^{+\nu}_{\tau}\partial_{\nu}\big(\rho^{\mu}_I\partial_{\mu}v\big)\xi^I\\
&\quad -\big(\rho^{\mu}_I\partial_{\mu}A_J-\rho^{\mu}_J \partial_{\mu}A_I-C^K_{IJ}A_K\big)\xi^I\psi_{\tau}^J+\frac{1}{2}\eta^{+\nu}_{\tau}\partial_{\nu}\big(\rho^{\mu}_I\partial_{\mu}A_J-\rho^{\mu}_J \partial_{\mu}A_I-C^K_{IJ}A_K\big)\xi^I\xi^J \\
&\quad +\sum_{n\geq 2}\left\{\frac{(-1)^{n}}{n!}\left(\sum_{k=1}^{n+1}(-1)^{k-1}\rho^{\mu}_{I_k}\partial_{\mu}\Omega^{(n)}_{I_1\cdots \widehat{I}_k\cdots I_{n+1}} \right. \right. \\
&\qquad\qquad \qquad\qquad\qquad \qquad\qquad\!\left.+ \sum_{1\leq k < l \leq n+1}(-1)^{k+l}C^{J}_{I_kI_l}\Omega^{(n)}_{JI_1\cdots \widehat{I}_k\cdots \widehat{I}_l\cdots I_{n+1}}\right) \xi^{I_1}\cdots\xi^{I_{n}}\psi^{I_{n+1}}_{\tau}\\
& \quad+\frac{(-1)^{n+1}}{(n+1)!}\eta^{+\nu}_{\tau}\partial_{\nu}\left(\sum_{k=1}^{n+1}(-1)^{k-1}\rho^{\mu}_{I_k}\partial_{\mu}\Omega^{(n)}_{I_1\cdots \widehat{I}_k\cdots I_{n+1}}\right.\\
&\qquad\qquad \qquad\qquad\qquad \qquad\qquad\!\!\left.\left.+ \sum_{1\leq k < l \leq n+1}(-1)^{k+l}C^{J}_{I_kI_l}\Omega^{(n)}_{JI_1\cdots \widehat{I}_k\cdots \widehat{I}_l\cdots I_{n+1}}\right)\xi^{I_1}\cdots\xi^{I_{n+1}}\right\}.
\end{split}
\end{align}
Throughout this calculation, use is made of the fact that $v$ and $\Omega^{(2k)}_{I_1\cdots I_{2k}}$ anticommute with $\eta^{+\mu}_{\tau}$ and $\xi^I$; this, together with the rules described in section~\ref{Sect:AKSZ}, determines the relevant signs. The counterpart of expression \eqref{eq:7.6}, on the other hand, is given by
\begin{align}
\label{eq:9.4}
\begin{split}
&U(0,\tau)^{-1}\partial_{\tau}\left(v+A_I \xi^I+\sum_{k\geq 2}\frac{1}{k!}\Omega^{(k)}_{I_1\cdots I_k}\xi^{I_1}\cdots\xi^{I_k} \right)U(0,\tau) \\
&\qquad \qquad=\partial_{\tau}\left\{U(0,\tau)^{-1}\left(v+A_I \xi^I+\sum_{k\geq 2}\frac{1}{k!}\Omega^{(k)}_{I_1\cdots I_k}\xi^{I_1}\cdots\xi^{I_k} \right)U(0,\tau)\right\} \\
&\qquad\qquad\quad+U(0,\tau)^{-1}\left[v+A_I \xi^I+\sum_{k\geq 2}\frac{1}{k!}\Omega^{(k)}_{I_1\cdots I_k}\xi^{I_1}\cdots\xi^{I_k} ,M \right]U(0,\tau),
\end{split}
\end{align}
where (as before) the bracket $[,]$ in the last term on the right-hand side denotes the usual commutator. This can be evaluated after a direct but tedious calculation, with the result
\begin{align}
\label{eq:9.5}
\begin{split}
&\left[v+A_I \xi^I+\sum_{k\geq 2}\frac{1}{k!}\Omega^{(k)}_{I_1\cdots I_k}\xi^{I_1}\cdots\xi^{I_k} ,M \right]\\
&\qquad=-\eta^{+\nu}_{\tau}\partial_{\nu}\big(v^2\big)-[A_I,v]\psi^I_{\tau}+\eta^{+\nu}_{\tau}\partial_{\nu}[A_I,v]\xi^I\\
&\qquad\quad+\big( v\Omega^{(2)}_{IJ}+\Omega^{(2)}_{IJ}v+[A_I,A_J]\big)\xi^I \psi^J_{\tau} -\frac{1}{2!}\eta^{+\nu}_{\tau}\partial_{\nu}\big(v\Omega^{(2)}_{IJ}+\Omega^{(2)}_{IJ}v+[A_I,A_J]\big)\xi^I \xi^J \\
&\qquad\quad- \sum_{n\geq 2}\left\{\frac{(-1)^{n}}{n!}\left( (-1)^{n+1}v \Omega^{(n+1)}_{I_1\cdots I_{n+1}}+\Omega^{(n+1)}_{I_1\cdots I_{n+1}}v+\sum_{k=1}^{n+1}(-1)^{k-1}\left[A_{I_k},\Omega^{(n)}_{I_1\cdots \widehat{I}_k\cdots I_{n+1}}\right]\right.\right.\\
&\qquad\quad \qquad\qquad\qquad\qquad\!\left.+\sum_{k=2}^{n-1}\binom{n+1}{k}(-1)^{(1-k)(n+1-k)}\Omega^{(k)}_{I_1\cdots I_k}\Omega^{(n+1-k)}_{I_{k+1}\cdots I_{n+1}}\right) \xi^{I_1}\cdots \xi^{I_n}\psi^{I_{n+1}}_{\tau}\\
&\quad\qquad+\frac{(-1)^{n+1}}{(n+1)!}\eta^{+\nu}_{\tau}\partial_{\nu}\left((-1)^{n+1}v \Omega^{(n+1)}_{I_1\cdots I_{n+1}}+\Omega^{(n+1)}_{I_1\cdots I_{n+1}}v+\sum_{k=1}^{n+1}(-1)^{k-1}\left[A_{I_k},\Omega^{(n)}_{I_1\cdots \widehat{I}_k\cdots I_{n+1}}\right] \right.\\
&\qquad\quad \qquad\qquad\qquad\qquad\!\!\left.\left.+\sum_{k=2}^{n-1}\binom{n+1}{k}(-1)^{(1-k)(n+1-k)}\Omega^{(k)}_{I_1\cdots I_k}\Omega^{(n+1-k)}_{I_{k+1}\cdots I_{n+1}}\right) \xi^{I_1}\cdots \xi^{I_{n+1}}\right\}.
\end{split}
\end{align}
Combining \eqref{eq:9.3} with \eqref{eq:9.4} and \eqref{eq:9.5} then yields
\begin{align}
\label{eq:9.6}
\begin{split}
&U(0,\tau)^{-1}\delta M(\tau) U(0,\tau) \\
&\qquad = -\partial_{\tau}\left\{U(0,\tau)^{-1}\left(v+A_I \xi^I+\sum_{k\geq 2}\frac{1}{k!}\Omega^{(k)}_{I_1\cdots I_k}\xi^{I_1}\cdots\xi^{I_k} \right)U(0,\tau)\right\}\\
&\qquad \quad+U(0,\tau)^{-1} \left( \eta^{+\nu}\partial_{\nu}\big(v^2\big)+\nabla_{I}v\psi^{I}_{\tau}-\eta^{+\nu}_{\tau}\partial_{\nu}\big(\nabla_{I}v\big)\psi^{I}_{\tau}\vphantom{\sum_{k=1}^{n+1}}\right. \\
&\qquad \quad- \big(v\Omega^{(2)}_{IJ}+\Omega^{(2)}_{IJ}v+F_{IJ}\big)\xi^I\psi_{\tau}^J +\frac{1}{2}\eta^{+\nu}_{\tau}\partial_{\nu}\big(v\Omega^{(2)}_{IJ}+\Omega^{(2)}_{IJ}v+F_{IJ}\big)\xi^I\xi^J \\
&\qquad \quad+ \sum_{n\geq 2}\left\{\frac{(-1)^{n}}{n!}\left( (-1)^{n+1}v \Omega^{(n+1)}_{I_1\cdots I_{n+1}}+\Omega^{(n+1)}_{I_1\cdots I_{n+1}}v + \big( \mathrm{d}_{F,\nabla}\Omega^{(n)} \big)_{I_1 \cdots I_{n+1}} \vphantom{\sum_{k=1}^{n+1}}\right.\right.\\
&\qquad \quad \!\left.+\sum_{k=2}^{n-1}\binom{n+1}{k}(-1)^{(1-k)(n+1-k)}\Omega^{(k)}_{I_1\cdots I_k}\Omega^{(n+1-k)}_{I_{k+1}\cdots I_{n+1}}\right) \xi^{I_1}\cdots \xi^{I_n}\psi^{I_{n+1}}_{\tau}\\
&\qquad \quad+\frac{(-1)^{n+1}}{(n+1)!}\eta^{+\nu}_{\tau}\partial_{\nu}\left((-1)^{n+1}v \Omega^{(n+1)}_{I_1\cdots I_{n+1}}+\Omega^{(n+1)}_{I_1\cdots I_{n+1}}v + \big( \mathrm{d}_{F,\nabla}\Omega^{(n)} \big)_{I_1 \cdots I_{n+1}} \vphantom{\sum_{k=1}^{n+1}}\right.\\
&\qquad \quad \!\!\!\left.\left.\left.+\sum_{k=2}^{n-1}\binom{n+1}{k}(-1)^{(1-k)(n+1-k)}\Omega^{(k)}_{I_1\cdots I_k}\Omega^{(n+1-k)}_{I_{k+1}\cdots I_{n+1}}\right) \xi^{I_1}\cdots \xi^{I_{n+1}}\right\}\right)U(0,\tau),
\end{split}
\end{align}
where $\mathrm{d}_{F,\nabla}$ is the exterior covariant derivative induced by the $F$-connection $\nabla$, and $F_{IJ}$ are the components of the $F$-curvature. (We have abbreviated $v(\phi^{\mu}(\tau))$ to $v$, $\Omega^{(2)}_{IJ}(\phi^{\mu}(\tau))$ to $\Omega^{(2)}_{IJ}$, and so on, to avoid a clash of notations.) Inserting this expression back into \eqref{eq:9.3}, we arrive at
\begin{align}
\label{eq:9.7}
\begin{split}
&\delta \str U(0,1)  = -\str\left\{U(0,1)\int_{0}^1U(0,\tau)^{-1} \left( \eta^{+\nu}\partial_{\nu}\big(v^2\big)+\nabla_{I}v\psi^{I}_{\tau}-\eta^{+\nu}_{\tau}\partial_{\nu}\big(\nabla_{I}v\big)\psi^{I}_{\tau}\vphantom{\sum_{k=1}^{n+1}}\right.\right. \\
&\qquad\quad\qquad\qquad\quad \quad- \big(v\Omega^{(2)}_{IJ}+\Omega^{(2)}_{IJ}v+F_{IJ}\big)\xi^I\psi_{\tau}^J +\frac{1}{2}\eta^{+\nu}_{\tau}\partial_{\nu}\big(v\Omega^{(2)}_{IJ}+\Omega^{(2)}_{IJ}v+F_{IJ}\big)\xi^I\xi^J \\
&\qquad\quad\qquad \qquad\quad \quad+ \sum_{n\geq 2}\left\{\frac{(-1)^{n}}{n!}\left( (-1)^{n+1}v \Omega^{(n+1)}_{I_1\cdots I_{n+1}}+\Omega^{(n+1)}_{I_1\cdots I_{n+1}}v + \big( \mathrm{d}_{F,\nabla}\Omega^{(n)} \big)_{I_1 \cdots I_{n+1}} \vphantom{\sum_{k=1}^{n+1}}\right.\right.\\
&\qquad\quad\qquad \qquad\quad\quad \!\left.+\sum_{k=2}^{n-1}\binom{n+1}{k}(-1)^{(1-k)(n+1-k)}\Omega^{(k)}_{I_1\cdots I_k}\Omega^{(n+1-k)}_{I_{k+1}\cdots I_{n+1}}\right) \xi^{I_1}\cdots \xi^{I_n}\psi^{I_{n+1}}_{\tau}\\
&\qquad\quad\qquad \qquad\quad \quad+\frac{(-1)^{n+1}}{(n+1)!}\eta^{+\nu}_{\tau}\partial_{\nu}\left((-1)^{n+1}v \Omega^{(n+1)}_{I_1\cdots I_{n+1}}+\Omega^{(n+1)}_{I_1\cdots I_{n+1}}v + \big( \mathrm{d}_{F,\nabla}\Omega^{(n)} \big)_{I_1 \cdots I_{n+1}}\vphantom{\sum_{k=1}^{n+1}}\right.\\
&\qquad\quad\qquad \qquad\quad \quad \!\!\!\!\left.\left.\left.\left.+\sum_{k=2}^{n-1}\binom{n+1}{k}(-1)^{(1-k)(n+1-k)}\Omega^{(k)}_{I_1\cdots I_k}\Omega^{(n+1-k)}_{I_{k+1}\cdots I_{n+1}}\right) \xi^{I_1}\cdots \xi^{I_{n+1}}\right\}\right)U(0,\tau)\d \tau\right\}.
\end{split}
\end{align}
In obtaining \eqref{eq:9.7}, we used the periodicity of $\phi^{\mu}$ and $\xi^I$, together with the fact that $U(0,1)$ carries zero ghost number, to show that the contribution of the total derivative term in \eqref{eq:9.6} vanishes:
\begin{align*}
&\str\left\{\left(v(\phi^{\mu}(1))+A_I(\phi^{\mu}(1)) \xi^I(1)+\sum_{k\geq 2}\frac{1}{k!}\Omega^{(k)}_{I_1\cdots I_k}(\phi^{\mu}(1))\xi^{I_1}(1)\cdots\xi^{I_k}(1) \right)U(0,1)\right\} \\
&\qquad\qquad-\str\left\{U(0,1)\left(v(\phi^{\mu}(0))+A_I(\phi^{\mu}(0)) \xi^I(0)+\sum_{k\geq 2}\frac{1}{k!}\Omega^{(k)}_{I_1\cdots I_k}(\phi^{\mu}(0))\xi^{I_1}(0)\cdots\xi^{I_k}(0) \right)\right\}=0.
\end{align*}
The condition that $\str U(0,1)$ be BRST-invariant may thus be formulated as
\begin{align*}
& v^2=0,\\
&\nabla_I v=0, \\
& v\Omega^{(2)}_{IJ}+\Omega^{(2)}_{IJ}v +F_{IJ}=0,
 \end{align*}
and 
\begin{align*}
(-1)^{n+1}v \Omega^{(n+1)}_{I_1\cdots I_{n+1}}+\Omega^{(n+1)}_{I_1\cdots I_{n+1}}v + \big( \mathrm{d}_{F,\nabla}\Omega^{(n)} \big)_{I_1 \cdots I_{n+1}}+\sum_{k=2}^{n-1}\binom{n+1}{k}(-1)^{(1-k)(n+1-k)}\Omega^{(k)}_{I_1\cdots I_k}\Omega^{(n+1-k)}_{I_{k+1}\cdots I_{n+1}} =0,
\end{align*}
for all $n \geq 2$. Referring back to equations \eqref{eq:localRepsGraded1}-\eqref{eq:localRepsGraded2} of section~\ref{Sect:2}, this means that the $F$-superconnection $D$ must in fact be a representation up to homotopy of $F$ on the $\ZZ$-graded vector bundle $V$.

\section{Boundary observables coupled to $F$-superconnections}
\label{Sect:10}
Finally, let us discuss the BRST-invariant boundary observables that can be defined after coupling to an $F$-superconnection. To this end, we may use a line of reasoning analogous to that of section~\ref{Sect:8}. All conventions and notations adopted in the preceding section are retained here as well. 

To begin, we define for any $\Hom(V^{\bullet},V^{\bullet+p})$-valued $F$-$k$-form $\omega=\frac{1}{k!}\omega_{I_1 \cdots I_k}e^{I_1}\wedge \cdots \wedge e^{I_k}$ the functional
$$
 \mathcal{O}_{\omega}=\frac{1}{k!}\omega_{I_1 \cdots I_k}(\gbf{\phi})\gbf{\psi}^{I_1}\cdots\gbf{\psi}^{I_k}.
$$
In components this becomes 
$$
\mathcal{O}_{\omega}= \mathcal{O}_{\omega}^{(0)}+\theta  \mathcal{O}_{\omega\tau}^{(1)},
$$
where $\mathcal{O}_{\omega}^{(0)}$ and $\mathcal{O}_{\omega \tau}^{(1)}$ are given by the same formulas as in \eqref{eq:6.3}. This being so, we have that under a BRST transformation these components change according to \eqref{eq:6.4}. We note in addition that $ \mathcal{O}_{\omega}^{(0)}$ and $\mathcal{O}_{\omega\tau}^{(1)}$ carry ghost numbers $k+p$ and $k-1+p$, respectively.

Now we repeat the same steps as before. Generalizing \eqref{eq:7.3}, we consider the path integral
\begin{equation}
\label{eq:10.1}
\int_{\mathfrak{L}} D\Phi \exp(-S) \cdot \str \left(H(t)\mathcal{O}_{\omega}^{(0)}(t)\right).
\end{equation}
Here as in the discussion of section~\ref{Sect:8}, $H(t)$ is the conjugacy class of the holonomy $U(0,1)$, and `$\str$' denotes the supertrace on $\End(V_{\Phi(j(t))}) \otimes \Lambda^{\bullet}F^*_{\Phi(j(t))}$. We want to determine the condition on $\mathcal{O}_{\omega}^{(0)}$ that ensures that \eqref{eq:10.1} is BRST-invariant. For this purpose, it will be convenient to work with the element $\alpha$ defined by equation \eqref{eq:2.6}:
$$
\alpha=v + A_I e^{I} + \sum_{n\geq 2}\frac{1}{n!}\Omega^{(n)}_{I_1\cdots I_n}e^{I_1}\wedge \cdots \wedge e^{I_n}.
$$
To it, we associate the functional
$$
\mathcal{A}_{\alpha}=v+A_I\xi^I+\sum_{n\geq 2}\frac{1}{n!}\Omega^{(n)}_{I_1\cdots I_n}\xi^{I_1} \cdots  \xi^{I_n}.
$$
Then the Maurer-Cartan equation \eqref{eq:2.7} may be rewritten in the equivalent form
\begin{equation}
\label{eq:10.2}
-\delta \mathcal{A}_{\alpha}+\frac{1}{2}\big[ \mathcal{A}_{\alpha},\mathcal{A}_{\alpha}\big]=0.
\end{equation}
Here and henceforth the bracket $[,]$ denotes the graded commutator.  

With these conventions, one may write the change in $H(t)$ under a BRST transformation as
\begin{equation*}
\delta H(t)=\big[\mathcal{A}_{\alpha}(t), H(t) \big].
\end{equation*}
From this we can readily calculate that
\begin{equation}
\label{eq:10.3}
\delta \str \left(H(t)\mathcal{O}_{\omega}^{(0)}(t)\right)=\str \left(H(t)\widehat{\delta}\mathcal{O}_{\omega}^{(0)}(t)\right),
\end{equation}
where we have introduced the modified BRST operator
\begin{equation}
\label{eq:10.4}
\widehat{\delta}\mathcal{O}_{\omega}^{(0)}=\delta\mathcal{O}_{\omega}^{(0)}-\big[ \mathcal{A}_{\alpha},\mathcal{O}_{\omega}^{(0)} \big].
\end{equation}
The condition for \eqref{eq:10.1} to be invariant with respect to arbitrary BRST transformations is then
\begin{equation}
\label{eq:10.5}
\widehat{\delta}\mathcal{O}_{\omega}^{(0)}=0.
\end{equation}

The next thing to notice is that the modified BRST operator $\widehat{\delta}$ is nilpotent, just as it was in our analysis in section~\ref{Sect:8}. Indeed, a straightforward calculation gives
\begin{equation*}
\widehat{\delta}^2\mathcal{O}_{\omega}^{(0)}=\left[ -\delta \mathcal{A}_{\alpha}+\frac{1}{2}\big[ \mathcal{A}_{\alpha},\mathcal{A}_{\alpha}\big], \mathcal{O}_{\omega}^{(0)} \right],
\end{equation*}
which vanishes by virtue of \eqref{eq:10.2}. Equation \eqref{eq:10.5} therefore states that the boundary observables that interest us belong to the cohomology of $\widehat{\delta}$.

Now let us use the first equation in \eqref{eq:6.4} to rewrite \eqref{eq:10.4} as
\begin{equation}
\label{eq:10.6}
\widehat{\delta}\mathcal{O}_{\omega}^{(0)}=\mathcal{O}_{Q_{F,D}\omega}^{(0)},
\end{equation}
where $Q_{F,D}$ is the homological vector field on $\Pi F$ given by $Q_{F,D}=Q_F-[\alpha,\,]$. This, combined with the second equation in \eqref{eq:6.4}, yields
\begin{equation}
\label{eq:10.7}
\widehat{\delta}\mathcal{O}_{\omega}^{(1)}=\d \mathcal{O}_{\omega}^{(0)}-\mathcal{O}_{Q_{F,D}\omega}^{(1)}.
\end{equation}
Equation \eqref{eq:10.6} shows that the boundary observables of the form $\mathcal{O}_{\omega}^{(0)}$ are one-to-one correspondence with elements of the cohomology of $Q_{F,D}$. The latter can be simply understood by noting that, in view of the discussion surrounding \eqref{eq:6.4}, $Q_{F,D}$ is identified with the differential $-\partial_{D,D}=-\mathrm{d}_F-[\alpha,\,]$ acting on the space of $\End(V)$-valued $F$-forms. The $Q_{F,D}$-cohomology is therefore isomorphic to
$$
H^{\bullet}_{\partial_{D,D}}(\Omega^{\bullet}(F,\End(V))),
$$
where we have borrowed the notation of section~\ref{Sect:2}. Adding all these ingredients together, we see that boundary observables of the form $\mathcal{O}_{\omega}^{(0)}$ are represented by elements of the endomorphism algebra of $(V,D)$ in the homotopy category of the DG-category of representations up to homotopy of $F$. 

Next, we observe that if $\omega$ is fixed representative of a $Q_{F,D}$-cohomology class, the equation \eqref{eq:10.7} simplifies to
\begin{equation}
\label{eq:10.8}
\widehat{\delta}\mathcal{O}_{\omega}^{(1)}=\d \mathcal{O}_{\omega}^{(0)}.
\end{equation}
In words, the descent equations are produced by acting with $\widehat{\delta}$ rather than $\delta$. Since \eqref{eq:10.8} asserts that the $\mathcal{O}_{\omega}^{(1)}$ is annihilated by $\widehat{\delta}$ up to an exact form, we have the $\widehat{\delta}$-invariant boundary observable
$$
\int_{\partial \Sigma} \mathcal{O}_{\omega}^{(1)}.
$$
As usual, we can interpret this observable as a deformation of the boundary interaction of the theory. Concretely, this means that the formula \eqref{eq:7.2} for path integral can be modified to
$$
\int_{\mathfrak{L}}D\Phi \exp(-S)\cdot\str P\!\exp\left(\epsilon\int_{\partial \Sigma} \mathcal{O}_{\omega}^{(1)}  \right),
$$ 
where we treat $\epsilon$ as a formal parameter. 

It seems appropriate to conclude by stating explicitly the following generalization of these conside\-rations. Suppose that the circle $\partial \Sigma$ is split into two segments $(0,t)$ and $(t,1)$ by the points $\tau=0$ and $\tau=t$. Suppose further that on the segment $(0,t)$ the boundary interaction is described by $(V_1,D_1)$, while on the segment $(t,1)$ it is described by $(V_2,D_2)$. Given a $\Hom(V_1^{\bullet},V_2^{\bullet+p})$-valued $F$-$k$-form $\omega_{1}$ and a $\Hom(V_2^{\bullet},V_1^{\bullet+q})$-valued $F$-$l$-form $\omega_{2}$ consider the following expression to be inserted in the path integral  \eqref{eq:7.2}:
\begin{equation}
\label{eq:10.9}
\str \left(  \mathcal{O}_{\omega_{1}}^{(0)}(t)U_1(0,t) \mathcal{O}_{\omega_{2}}^{(0)}(0)U_2(0,t)^{-1} \right).
\end{equation}
Here the parallel transport operators $U_1(0,t)$ and $U_2(0,t)$ are constructed from $(V_1,D_1)$ and $(V_2,D_2)$, respectively. By calculations similar to those which led to \eqref{eq:10.3}, it can be seen that the change in \eqref{eq:10.9} under a BRST transformation is explicitly given by
\begin{align*}
&\delta \str \left(\mathcal{O}_{\omega_{1}}^{(0)}(t)U_1(0,t)\mathcal{O}_{\omega_{2}}^{(0)}(0)U_2(0,t)^{-1}\right) \\
&\qquad=\str \left(\widehat{\delta}_{D_1,D_2}\mathcal{O}_{\omega_{1}}^{(0)}(t)U_1(0,t)\mathcal{O}_{\omega_{2}}^{(0)}(0)U_2(0,t)^{-1}\right)+(-1)^{k+p}\str \left(\mathcal{O}_{\omega_{1}}^{(0)}(t)U_1(0,t)\widehat{\delta}_{D_2,D_1}\mathcal{O}_{\omega_{2}}^{(0)}(0)U_2(0,t)^{-1}\right),
\end{align*}
where we defined
\begin{align*}
\widehat{\delta}_{D_1,D_2}\mathcal{O}_{\omega_{1}}^{(0)}&=\delta\mathcal{O}_{\omega_{1}}^{(0)}-\mathcal{A}_{\alpha_2}\mathcal{O}^{(0)}_{\omega_{1}}+(-1)^{k+p}\mathcal{O}_{\omega_{1}}^{(0)} \mathcal{A}_{\alpha_{1}}, \\
\widehat{\delta}_{D_2,D_1}\mathcal{O}_{\omega_{2}}^{(0)}&=\delta\mathcal{O}_{\omega_{2}}^{(0)}-\mathcal{A}_{\alpha_1}\mathcal{O}^{(0)}_{\omega_{2}}+(-1)^{l+q}\mathcal{O}_{\omega_{2}}^{(0)} \mathcal{A}_{\alpha_2}.
\end{align*}
The condition for the BRST-invariance of \eqref{eq:10.9} then implies separately that $\widehat{\delta}_{D_1,D_2}\mathcal{O}_{\omega_{1}}^{(0)}=0$ and $\widehat{\delta}_{D_2,D_1}\mathcal{O}_{\omega_{2}}^{(0)}=0$. In like manner, it may be shown that $\widehat{\delta}_{D_1,D_2}$ and $\widehat{\delta}_{D_2,D_1}$ are both nilpotent. Thus, to find out the nontrivial boundary observables we are led to study the cohomology groups defined by $\widehat{\delta}_{D_1,D_2}$ and $\widehat{\delta}_{D_2,D_1}$. This analysis is identical to the one presented before. Applied to this case one finds that these cohomology groups are isomorphic to their counterparts
$$
H^{\bullet}_{\partial_{D_1,D_2}}(\Omega^{\bullet}(F,\Hom(V_1,V_2)))
$$
and
$$
H^{\bullet}_{\partial_{D_2,D_1}}(\Omega^{\bullet}(F,\Hom(V_2,V_1)))
$$
where we have again employed the notation of section~\ref{Sect:2}. Therefore we may regard morphisms from $(V_1,D_1)$ to $(V_2,D_2)$ and from $(V_2,D_2)$ to $(V_1,D_1)$ as boundary observables to be inserted at the joining point of boundary conditions labeled by $(V_1,D_1)$ and $(V_2,D_2)$. We will  loosely refer to these as boundary-changing observables.

\section{Conclusions and outlook regarding topological D-branes}
\label{Sect:11}
In this paper, we have studied the boundary coupling of a Lie algebroid Poisson sigma model with an arbitrary Lie algebroid $E \to X$ as target. We have seen that, for a boundary condition described by a Lie subalgebroid $F$ of $E$ over a submanifold $Y \subset X$, such a model can be coupled to a representation up to homotopy $(V,D)$ of $F$ while preserving the BRST symmetry. Anticipating the application to open topological string theory, we may refer to $(V,D)$ as a D-brane wrapped on $Y$. Given two such D-branes $(V,D)$ and $(V',D')$, we have shown that the space of boundary-changing observables can be identified with the space of morphisms from $(V,D)$ to $(V',D')$ in the homotopy category of the DG-category of representations up to homotopy of $F$, which we denoted by $\Ho\Rep(F)$ in section~\ref{Sect:2}. This can be rephrased more generally saying that the set of D-branes wrapping $Y$ has the structure of a linear category (in fact triangulated), with morphisms corresponding to boundary-changing observables. Let us comment on the likely implications of these results for the geometry of D-branes in topologically twisted nonlinear sigma models.

It has been argued in \cite{Kap04,KL05,KL07} that the geometry of D-branes in general topologically twisted nonlinear sigma models can be conveniently described in the language of generalized complex structures as introduced by Hitchin \cite{Hitchin03}. To set the stage, let us briefly recall the definition of the latter. A {\it generalized complex structure} on a smooth manifold $X$ is a bundle map $\mathcal{J}\colon TX\oplus T^*X \to TX\oplus T^*X$ that preserves the obvious pseudo-Euclidean metric on $TX\oplus T^*X$, satisfies $\mathcal{J}^2=-1$, and whose $+i$-eigenbundle $E \subset (TX \oplus T^*X)\otimes \CC$ is closed with respect to the so-called Courant bracket. (For detail, see Gualtieri's thesis \cite{Gualt03}.) Compatibility with the pseudo-Euclidean metric implies that $E$ is a maximal isotropic subbundle of $(TX \oplus T^*X)\otimes \CC$, and thus the Courant bracket on $TX \oplus T^*X$ endows $E$ with the structure of a complex Lie algebroid.\footnote{A complex Lie algebroid is the same as a Lie algebroid, except that $E$ is a complex vector bundle, and $TX$ is replaced with its complexification $TX\otimes \CC$.} Since $E$ determines $\mathcal{J}$ completely, we see that a generalized complex structure may be equivalently defined as a complex Lie algebroid $E$ with the additional property that $E^*=\overline{E}$.  

The simplest example of a generalized complex structure is
$$
\mathcal{J}_{J}=\left( \begin{matrix} J & 0 \\ 0 & -J^{*} \end{matrix}\right),
$$ 
where $J\colon TX \to TX$ is an ordinary complex structure, and $J^{*}\colon T^*X\to T^*X$ is its dual. The $+i$-eigenbundle $E_J$ of $\mathcal{J}_J$ is equal to $TX^{0,1}\oplus T^*X^{1,0}$, where $TX^{1,0}$ is the $+i$-eigenbundle of $J$ in the usual way. If $\omega$ is a symplectic structure, then a second example of generalized complex structure is given by
$$
\mathcal{J}_{\omega}=\left( \begin{matrix} 0 & -\omega^{-1} \\ \omega & 0 \end{matrix}\right).
$$ 
Here we regard $\omega$ as a bundle map from $TX$ to $T^*X$; $\omega^{-1}$ is the inverse map from $T^*X$ to $TX$. The $+i$-eigenbundle $E_{\omega}$ of $\mathcal{J}_{\omega}$ is the graph of $-i \omega$, and hence is isomorphic to $TX \otimes \CC$ as a complex Lie algebroid. 

The above discussion makes it clear that to any generalized complex structure $\mathcal{J}$ on a manifold $X$ we can associate a Lie algebroid Poisson sigma model with target geometry specified by the corresponding complex Lie algebroid $E$. The key point, however, is that for a generalized complex structure $\mathcal{J}_J$ coming from a complex structure $J$ on $X$, the model in question is, upon gauge fixing, equivalent to the B-model on $X$; similarly, when $\mathcal{J}_{\omega}$ comes from a symplectic structure $\omega$ on $X$, the model is equivalent to the A-model on $X$ (see the remarks at the end of section~6 of \cite{CQZ10}). Thus, the results of this paper lead us to expect that the geometry of A- and B-branes can be completely rephrased in terms of representations up to homotopy. More generally, they suggest that given any generalized complex structure $\mathcal{J}$ on $X$ one should be able to define D-branes and morphisms between them. Let us proceed to explain and justify these statements.

First of all, we need to bring in the notion of generalized complex submanifold of a generalized complex manifold $X$. Let the {\it generalized tangent bundle} $\mathcal{T}_{Y,B}$ of a submanifold $Y$ of $X$, carrying a closed two-form $B$, be the subbundle of $(TX \oplus T^*X)\vert_{Y}$ determined by the condition that $v+\xi$ belong to $\mathcal{T}_{Y,B}$ if and only if $v$ belongs to $TY$ and the image of $\xi$ under the projection to $T^*Y$ is equal to the interior product $i_vB$. A {\it generalized complex submanifold} of $X$ is defined to be a pair $(Y,B)$ such that its generalized tangent bundle $\mathcal{T}_{Y,B}$ is stable under the action of $\mathcal{J}$. 

If $(Y,B)$ is a generalized complex submanifold of $X$, we let $E_Y$ denote the $+i$-eigenbundle of the restriction of $\mathcal{J}$ to $\mathcal{T}_{Y,B}$. This is a subbundle of the complexification of  $\mathcal{T}_{Y,B}$. What is more, there is a natural complex Lie algebroid structure on $E_Y$. The anchor map is the obvious projection to $TY \otimes \CC$. The Lie bracket is given by extending sections off $Y$, taking their Courant bracket, and restricting back to $Y$. One can easily check that the result lies in $E_Y$ and does not depend on how we extend sections off $Y$; see \cite{Fer02,Gualt03,KL05}. Finally, it is trivial from the pertinent definitions that $E_Y$ can be interpreted as a Lie subalgebroid of $E$.

In the case when $\mathcal{J}_J$ comes from a complex structure $J$ on $X$, a generalized complex submanifold is simply a complex submanifold $Y$ with a closed two-form $B$ of type $(1,1)$. For such a submanifold, $E_{J,Y}$ is isomorphic as a complex Lie algebroid to $TY^{0,1}\oplus N^*Y^{1,0}$.  

If $\mathcal{J}_{\omega}$ comes from a symplectic form $\omega$ on $X$, a generalized complex submanifold corresponds to a coisotropic submanifold $Y$ with an additional structure: if we denote by $\mathcal{L}Y$ the the kernel of $\omega\vert_Y$ and by $\mathcal{F}Y$ the quotient bundle $TY/\mathcal{L}Y$, then $\mathcal{L}Y$ induces a foliation of $Y$, $B$ descends to a section of $\Lambda^2 \mathcal{F}Y^*$, and $J=\omega^{-1}B\vert_{\mathcal{F}Y}$ defines a transverse complex structure on $Y$. It is easy to see that both $B$ and $\omega\vert_{\mathcal{F}Y}$ are of type $(2,0)+(0,2)$ with respect to this transverse complex structure; furthermore, the complex form $B+i\omega\vert_{\mathcal{F}Y}$ is of type $(2,0)$ and so it defines a holomorphic symplectic structure on the leaves. For such coisotropic submanifolds, the corresponding complex Lie algebroid $E_{\omega,Y}$ is isomorphic to $(\mathcal{L}Y\otimes \CC)\oplus \mathcal{F}Y^{1,0}$. Note finally that if $B=0$, then $Y$ is simply a Lagrangian submanifold, and $E_{\omega,Y}$ is isomorphic to $TY\otimes \CC$ as a complex Lie algebroid. 

The lesson of the foregoing discussion is that generalized complex submanifolds give rise to the correct boundary conditions for the Lie algebroid Poisson sigma model associated to a generalized complex manifold $X$. More fundamentally, as stated at the beginning of this section, D-branes wrapped on a generalized complex submanifold $(Y,B)$ correspond to representations up to homotopy of the complex Lie algebroid $E_Y$. This generalizes an earlier proposal of Gualtieri \cite{Gualt07,Gualt10}, in which D-branes supported on $(Y,B)$ are regarded as ordinary representations of $E_Y$. 

We will now argue that if $\mathcal{J}_J$ comes from a complex structure $J$ on $X$, this definition is in accordance with the more conventional definition of B-brane. The first relevant observation is that, as far as the B-model is concerned, it is sufficient to consider D-branes supported on the whole of $X$, henceforth referred to as ``space-filling D-branes''. (For an explanation of this point, see sections~3.6.3 and 5.3.3 of \cite{ABCDGKMSSW09}.) Also note that, in the case at hand, a generalized complex submanifold with $Y=X$ is specified by the choice of a closed two-form $B$ of type $(1,1)$; moreover, as already pointed out, $E_{J,X}=TX^{0,1}$. Given the way we have set things up, it is a straightforward matter to check that a space-filling D-brane consists of a $\ZZ$-graded smooth vector bundle $V=\bigoplus_{k\in \ZZ}V^k$ equipped with a flat $\overline{\partial}$-superconnection $D$, i.e.~a degree $1$ operator on the space of $(0,\bullet)$-forms with values in $V$ that satisfies a graded Leibniz rule and squares to zero. A theorem of Block \cite{Block10} tells us that this data is equivalent to that of an object in $\mathbf{D}^b(\coh(X))$, the bounded derived category of coherent sheaves on $X$. In fact, even more is true, namely that the homotopy category whose objects are given by these superconnections, which in our present terminology is simply $\Ho\Rep(TX^{0,1})$, is equivalent to $\mathbf{D}^b(\coh(X))$. The latter is widely believed to be equivalent to the category of B-branes on $X$. In parti\-cular, the space of boundary-changing observables is identified with the space of open string states between B-branes (see \cite{KS02} for more detail). It should perhaps be remarked that a non-vanishing two-form $B$ has the effect of tensoring all B-branes by a Hermitian line bundle over $X$ with a connection with curvature $B$. This clearly has no effect on the spaces of open strings, and therefore leaves the category of B-branes unchanged.

In the symplectic case, where $\mathcal{J}_{\omega}$ comes from a symplectic form $\omega$, things are more complicated. According to our general discussion, a D-brane in this context must be supported on a coisotropic submanifold $Y$ with a transverse complex structure $J=\omega^{-1}B\vert_{\mathcal{F}Y}$. When such a D-brane is labeled by an ordinary representation of the Lie algebroid $E_{\omega,Y}=(\mathcal{L}Y\otimes \CC)\oplus \mathcal{F}Y^{1,0}$, we obtain precisely the notion of an A-brane as a transversely holomorphic vector bundle over $Y$, i.e., a smooth vector bundle $V$ endowed with a connection whose curvature is of type $(1,1)$ with respect to the transverse complex structure $J=\omega^{-1}B\vert_{\mathcal{F}Y}$ and vanishes along the leaves of the foliation defined by $\mathcal{L}Y$; see \cite{Gualt07,Gualt10}. With this identification understood, it is not difficult to convince oneself that a D-brane corres\-ponding to a representation up to homotopy of $E_{\omega,Y}$ is the same thing as a complex of transversely holomorphic vector bundles over $Y$ with transversely holomorphic differentials (see e.g.~\cite{Gomez-Mont80} for a definition and discussion).\footnote{It is, perhaps, worth mentioning that the $\ZZ$-grading is provided by the axial R-charge (if it is not anomalous); cf.~\cite{Li06} for more information.} There is, however, one tricky point that we have glossed over. In addition to being transversely holomorphic, the coisotropic submanifold $Y$ also carries a transverse holomorphic symplectic form $\sigma=B+i\omega\vert_{\mathcal{F}Y}$. Recently, it was argued by Herbst \cite{Herbst10} that, in the A-model, the ring of BRST-invariant observables on $Y$, i.e., the ring of smooth functions on $Y$ which are locally constant along the leaves and holomorphic in the transverse directions, is deformed via deformation quantization using the Poisson brackets derived from the holomorphic symplectic form~$\sigma$. Taking this into account, Herbst then went on to show that an A-brane on $Y$ is determined by a complex of {\it noncommutative} transversely holomorphic vector bundles over $Y$ with {\it noncommutative} transversely holomorphic differentials (that is, the condition to be a differential involves the star-product; see section~3.2 of \cite{Herbst10} for details). This would seem to imply that we should reconsider our construction of the boundary coupling in the A-model so as to include representations up to homotopy over the quantized ring of BRST-invariant observables on $Y$. We hope to return to this issue in future work.

Another question which arises in this context is the following. So far we have considered only the case of a D-brane wrapped on a single coisotropic submanifold $Y$ with a fixed transverse complex structure $J$. However, more generally we would like to describe collections of D-branes supported on different coisotropic submanifolds with different transverse complex structures. For this, we must understand the space of boundary-changing observables between D-branes wrapped on two coisotropic submanifolds $Y_1$ and $Y_2$, with transverse complex structures $J_1$ and $J_2$, respectively. Once this is done, we will have a natural structure of a category on the set of D-branes.
Combined with our previous observation on the noncommutative deformation of the D-brane geometry, this would provide a new definition of the category of A-branes and it might help in the construction of an algebraic model of the Fukaya category. We expect to report on this in the future.

\begin{acknowledgements}
I wish to thank Camilo Arias Abad, Alberto Cattaneo and Manfred Herbst for helpful comments and useful correspondences. I am also grateful to Alastair Craw for his suggestions and advice.  Finally, I am indebted to the two anonymous referees for their careful reading, corrections, and suggestions. The author is supported by EPSRC grant EP/G004048.
\end{acknowledgements}


\addcontentsline{toc}{section}{Bibliography}
\end{document}